\documentclass[a4paper,11pt]{article}
\usepackage{jcappub}
\usepackage{bm}
\def\Z#1{_{\lower2pt\hbox{$\scriptstyle#1$}}}
\def\X#1{_{\lower1pt\hbox{$\scriptscriptstyle#1$}}}
\def\ns#1{_{\rm #1}} \def\Ns#1{_{\lower2pt\hbox{$\scriptstyle\rm#1$}}}
\def\dd{{\rm d}}   \def\etal{{et al}.}
\def\be{\beta}\def\ga{\gamma}\def\de{\delta}
\def\th{\theta}\def\rh{\rho}\def\si{\sigma}
\def\goesas{\mathop{\sim}\limits}\def\w#1{\,\hbox{#1}}
\def\lsim{\mathop{\hbox{${\lower3.8pt\hbox{$<$}}\atop{\raise0.2pt\hbox{$
\sim$}}$}}}
\def\gsim{\mathop{\hbox{${\lower3.8pt\hbox{$>$}}\atop{\raise0.2pt\hbox{$
\sim$}}$}}} \def\sith{\si\Z\th} \def\siz{\si_z}
\def\h{\,h^{-1}}\def\hm{\h\hbox{Mpc}}\def\deg{^\circ}
\def\vo{v_{\rm o}}\def\bo{\beta_{\rm o}}\def\s{\!\vphantom{H}}
\def\lcmb{\ell_{\rm o}}\def\bcmb{b_{\rm o}}\def\LCDM{$\Lambda$CDM}
\def\HO{H\Z0}\def\aO{a\Z0}\def\qO{q\Z0}\def\jO{j\Z0}\def\deO{\de\Z0}
\def\rO{r\Z0}\def\tO{t\Z0}\def\TO{T\Z0}
\def\robs{r\ns{obs}}
\def\kms{\w{km}\w{s}^{-1}}\def\kmsMpc{\kms\w{Mpc}^{-1}}
 \def\Ha{\bar H\Z0}
\def\ave#1{\left\langle#1\right\rangle}
\def\PRL#1{{\em Phys.\ Rev.\ Lett}.\ {\bf#1}}
\def\PR#1#2{{\em Phys.\ Rev.}\ {\bf#1 #2}}
\def\JCAP#1{{\em JCAP} {\bf#1}}
\def\ApJ#1{{\em Astrophys.\ J}.\ {\bf#1}}
\def\AaA#1{{\em Astron.\ Astrophys}.\ {\bf#1}}
\def\AJ#1{{\em Astron.\ J}.\ {\bf#1}}
\def\MNRAS#1{{\em Mon.\ Not.\ Roy.\ Astron.\ Soc}.\ {\bf#1}}
\def\CQG#1{{\em Class.\ Quant.\ Grav}.\ {\bf#1}}
\def\GRG#1{{\em Gen.\ Rel.\ Grav}.\ {\bf#1}}
\def\ApJs#1{{\em Astrophys.\ J.\ Suppl}.\ {\bf#1}}
\def\PLB#1{{\em Phys.\ Lett.}\ {\bf B #1}}
\def\beq{\begin{equation}} \def\eeq{\end{equation}}
\def\bea{\begin{eqnarray}} \def\eea{\end{eqnarray}}
\def\CS{{\textsc COMPOSITE}\ }
\def\wdi{w_{d,i}}\def\wzi{w_{z,i}}\def\wti{w_{\th,i}}\def\thi{\th_i}
\def\Tmod{T\Ns{AIE}}\def\zmod{z\Ns{AIE}}
\def\gmod{\gamma\Ns{AIE}}\def\bmod{{\bm\beta}\Ns{AIE}}
\def\gfl{\gamma\Ns{CMB}}\def\bfl{{\bm\beta}\Ns{CMB}}
\def\vmod{{\bm v}\Ns{AIE}}\def\vfl{{\bm v}\Ns{CMB}}
\def\nLI{\hat{\bm n}\Ns{AIE}}\def\nLG{\hat{\bm n}\Ns{LG}}
\def\nhel{\hat{\bm n}\Ns{hel}}
\newcommand{\av}[1]{\langle{#1}\rangle}
\providecommand{\href}[2]{#2}\def\link#1{\href{http://arxiv.org/abs/#1}{{\tt #1}}}
\title{Differential cosmic expansion and the Hubble flow anisotropy}
\author[a]{Krzysztof~Bolejko,}
\author[b]{M.~Ahsan Nazer}
\author[b]{and David~L.~Wiltshire}
\affiliation[a]{Sydney Institute for Astronomy, School of Physics, A28,
The University of Sydney, NSW 2006, Australia}
\affiliation[b]{Department of Physics and Astronomy,
University of Canterbury, Private Bag 4800, Christchurch 8140, New Zealand}
\emailAdd{bolejko@physics.usyd.edu.au}
\emailAdd{ahsan.nazer@canterbury.ac.nz}
\emailAdd{david.wiltshire@canterbury.ac.nz}
\arxivnumber{1512.07364}
\abstract{The Universe on scales $10$--$100\hm$ is dominated by
a cosmic web of voids, filaments, sheets and knots of galaxy clusters.
These structures participate differently in the global expansion of the
Universe: from non-expanding clusters to the above average expansion rate
of voids. In this paper we characterize Hubble expansion anisotropies in the
COMPOSITE sample of 4534 galaxies and clusters. We concentrate on the dipole and
quadrupole in the rest frame of the Local Group. These both have statistically
significant amplitudes. These anisotropies, and their redshift dependence,
cannot be explained solely by a boost of the Local Group in the
Friedmann-Lema\^{\i}tre-Robertson-Walker (FLRW) model which expands
isotropically in the rest frame of the cosmic microwave background (CMB)
radiation. We simulate the local expansion of the Universe with inhomogeneous
Szekeres solutions, which match the standard FLRW model on $\gsim100\hm$ scales
but exhibit nonkinematic relativistic differential expansion on small scales.
We restrict models to be consistent with observed CMB temperature anisotropies,
while simultaneously fitting the
redshift variation of the Hubble expansion dipole. We include features to
account for both the Local Void and the ``Great Attractor''. While this
naturally accounts for the Hubble expansion and CMB dipoles, the simulated
quadrupoles are smaller than observed. Further refinement to incorporate
additional structures may improve this. This would enable a test of the
hypothesis that some large angle CMB anomalies result from failing to treat
the relativistic differential expansion of the background geometry; a natural
feature of solutions to Einstein's equations not included in the current
standard model of cosmology.}
\begin{document}
\maketitle
\flushbottom
\section{Introduction}

In cosmology deviations from a uniform expansion
are most commonly treated as peculiar velocities relative to a linear
Hubble law
\beq
v\ns{pec}=cz-\HO r\,\label{vpec}
\eeq
where $z$ is the redshift, $c$ the speed of light, $r$ an appropriate distance
measure, and $\HO\equiv H(\tO)=100\,h\kmsMpc$ is the Hubble constant, $h$
being a dimensionless number. Such a theoretical framework is a natural
description if the assumption of homogeneity and isotropy holds at all
scales, so that all cosmologically relevant motions can be understood in
terms the background expansion of one single
Friedmann--Lema\^{\i}tre--Robertson--Walker (FLRW) geometry, with a
Hubble parameter, $H(t)$, given by the Friedmann equation, plus local boosts
which can be treated by eq.~(\ref{vpec}) for suitably small values of the
distance,\footnote{Even within FLRW models, for large values of $r$ one has to take into account that $H(t)$
varies with time, and that the redshift is not additive but rather a multiplicative quantity,
so a simple addition as in (\ref{vpec}) does not apply:
$1+z_{1+2} = (1+z_1) (1+z_2) \ne 1 + z_1 + z_2$.} $r$.

However, on scales of tens of megaparsecs the Universe
exhibits strong inhomogeneities, dominated in volume by voids with
density contrasts close to the minimum possible $\de\rh/\rh=-1$
\cite{hv02,hv04,minivoids,pan11}. Galaxies and galaxy clusters are not
randomly distributed but are strung in filaments that thread and surround
the voids to form a complex cosmic web \cite{web1,web2,web3}.
The Universe is only spatially homogeneous is some statistical
sense when one averages on scales $\gsim100\hm$. Just
how large this scale is, is debated \cite{hogg05,sl09,sdb12,chr,rbof14}.
However, based on the fractal dimension of the 2--point galaxy correlation
function making a gradual transition to the homogeneous limit $D_2\to3$
in three spatial dimensions, a scale of statistical homogeneity in the
range $70\lsim r\ns{ssh}\lsim120\hm$ seems to be observed \cite{sdb12}.

Despite the fact that the FLRW geometry can only be observationally justified
on $\gsim100\hm$ scales, by tradition it is conventionally assumed that
such a geometry is still applicable at all scales on which space is expanding
below $r\ns{ssh}\goesas100\hm$, that is, until one gets to the very small
scales of bound clusters of galaxies. However, this assumption is not justified
by the principles of general relativity. In general, in solutions of Einstein's
equations the background space does not expand rigidly to maintain constant
spatial curvature as it does in the FLRW geometry.
General inhomogeneous cosmological models, such as the
Lema\^{\i}tre--Tolman (LT) \cite{L,T,B} and Szekeres \cite{Sz} models,
exhibit differential cosmic expansion. The Hubble parameter becomes a
function of space as well as time, and any relation
(\ref{vpec}) can no longer have the physical sense of defining a peculiar
velocity field with respect to a single expansion rate.

In this paper we will present the results of numerical investigations that
quantify the nonlinearity associated with differential cosmic expansion
in Szekeres solutions chosen to match key features of both the Cosmic Microwave
Background (CMB) anisotropies, and also the Hubble expansion on $\lsim100\hm$
scales. The crucial feature in these simulations is that the dipole
induced by local inhomogeneities cannot be directly attributed to
a $635\kms$ local boost of the Local Group (LG) of galaxies, and is thus
nonkinematic (as we define more precisely in Section \ref{nonk} below).

Despite its naturalness in general relativity, the hypothesis of a
nonkinematic origin for a fraction of the CMB dipole
goes against the consensus of what has been assumed in observational
cosmology \cite{ss67,pw68} ever since the first bounds were placed on the
anisotropy of the CMB in the 1960s \cite{pw67}. By now it is standard
practice to automatically transform redshift data to the CMB rest frame
before performing cosmological analyses.

Within the standard peculiar velocities framework, the amplitude of
bulk flows, their consistency with the standard Lambda Cold Dark Matter
(\LCDM) cosmology and their convergence to the CMB frame, are matters of
ongoing debate
\cite*{wfh09,ltmc,kash10,cmss,dn11,turn12,lah12,ms13,minrep,Ppec,HCT,ctlm}.
A possible nonkinematic origin for a fraction of the CMB dipole would
impact directly on this debate, as well as suggesting a reexamination
of other observational puzzles. Arguably the most important of these are
the large angle anomalies that have been observed in the CMB anisotropy spectrum for over a decade \cite{toh03,ehb04,dOC04,sshc04,lm05,chss06,ebg07,heb09,kn10,afse14}, with a statistical significance that has increased with the release of Planck satellite data \cite{Piso,Piso2,Srev}.

The hypothesis that a fraction of the local Hubble expansion is nonkinematic
should be subject to appropriate
observational tests. In recent work \cite{rest,boost} we devised such tests
and found very strong Bayesian evidence for the nonkinematic hypothesis.
In an independent study \cite{rs13}, the hypothesis of a purely kinematic
origin for the dipole in the cosmic distribution of radio galaxies has been
rejected at the 99.5\% confidence level.

In order to develop more powerful tests of the nonkinematic differential
expansion hypothesis, in this paper we will use exact solutions of Einstein's
equations for structures smaller than the statistical homogeneity scale
\cite{sdb12} for the purpose of ray--tracing simulations.

There have been a number of previous studies which have used the LT solution
to model the effects of anisotropic expansion
\cite{afms93,afs94,hmm97,T00,aa06,bh08,antil,dopp}, including its effects
on the CMB. However, these have typically
considered the effects of voids at large distances from our location,
or the effects of voids much larger than the small scale
inhomogeneities we will consider.

To our knowledge this paper contains the first ever study which seeks
to use exact solutions of Einstein's equations to model structures giving
rise to nonlinear expansion on scales comparable to those observed, constrained
directly by both ray--tracing of the CMB and by the Hubble expansion field
from actual surveys. While the Szekeres solution has been employed for a
number of cosmological problems
\cite{B09,bc10,nit11,ip11,BS11,BS13,kh15a,kh15b,SG15a,SG15b}, we believe that
this is also the first time that it has been used for ray--tracing simulations
of local structures. We will see that although we are not able to match
all features of the nonlinear Hubble expansion below the statistical
homogeneity scale, with the Szekeres solution we can nonetheless match more
features of the actual data than with other models, including the standard
FLRW cosmology with a local boost of the Local Group of galaxies.

This paper provides a proof--of--principle demonstration that we hope
will encourage even more sophisticated investigations of relativistic effects
beyond the perturbed FLRW model. Some potential future investigations
are outlined in Sec.\ \ref{anomalies}.

\section{Terminology}

In this paper we often use terms such as {\em nonlinear} and {\em nonkinematic}. These terms are ambiguous and therefore this Section describes how these terms are defined in this paper.

\subsection{Nonkinematic and relativistic differential expansion}\label{nonk}
The fact that cosmic expansion can vary not only in time but in space leads to a variation in the redshift of observed astronomical sources, be they galaxies or the CMB. If the redshift of an observed object can be described solely in terms of a homogeneous expansion and a Doppler effect, then we would call the redshift anisotropy {\em kinematic}:
\beq
(1+z)\ns{obs} = (1+z)\Ns{FLRW} (1+z)\Ns{Doppler}\,.\label{kin}
\eeq
The first redshift term on the right hand side refers to the global homogeneous and isotropic expansion of the FLRW model,
and second term is due to a Doppler effect with respect to the FLRW background that combines the motions of the observer (local boost) and the observed object (peculiar motion).
If the redshift of an observed source (galaxies or CMB) cannot be explained entirely in terms of the above equation then {\em nonkinematic} effects are present, be they a real physical phenomenon or merely some observational bias\footnote{We assume
all observational biases can be accounted for, although this requires
care in actual data analysis \cite{boost}. We deal only with the case of real physical
effects in this paper. Furthermore, while all galaxies within larger bound
clusters will still be assumed to exhibit peculiar local motions within each
cluster, for nonkinematic redshift anisotropies we are only interested in
redshifts and distances assigned collectively to gravitationally
bound structures.}.

The factor $(1+z)\Ns{FLRW}$ in (\ref{kin}) can only be defined with
respect to a canonical choice of our local Lorentz frame. Since the CMB is
remarkably isotropic, with a dipole of amplitude $1.23\times10^{-3}\TO$
of the mean temperature $\TO=2.725\,$K, the canonical {\em CMB rest frame}
is defined by matching the $3.37\,$mK temperature dipole
to the dipole in the series expansion
\beq
{\TO\over\gfl(1-\beta\Ns{CMB}\cos\th)}=\TO\left[1+\beta\Ns{CMB}\cos\th+
\beta\Ns{CMB}^2\left(\cos^2\th-\frac12\right)+\dots\right]
\label{dipk}
\eeq
where $\beta\Ns{CMB}\cos\th\equiv\bfl\cdot\nhel$, $\nhel$ is the unit vector
on the sky in the heliocentric frame, $\bfl=\vfl/c$ is the boost vector of the
CMB frame in the heliocentric frame, and $\gfl=(1-\be\Ns{CMB}^2)^{-1/2}$ is the
standard Lorentz gamma factor.

Using measurements of redshifts and distances of galaxy clusters, one can
independently define the local {\em average isotropic expansion (AIE)
frame} as the Lorentz frame at our location in which the spherically
averaged distance--redshift relation in independent radial shells has minimal
variations relative to a linear Hubble law \cite{rest,boost}. Since isotropy
is only defined by an average then the observed redshift of any individual
source will in general display a nonkinematic anisotropy differing from
(\ref{kin}) with a dependence, $z(\nLI)$, on the unit vector on the sky in the
AIE frame, $\nLI$.

According to (\ref{kin}), even a perturbed FLRW model will display nonkinematic
effects. However, such effects are small and are not expected to affect the
identification of the dipole in (\ref{dipk}). In this paper, we will study
models with nonkinematic effects that we will characterize in terms of the
variation of averages of the nearby Hubble expansion. These nonkinematic effects
will turn out to be so large that they are also likely to be distinguishable
when compared to perturbed FLRW models, offering a simple alternative
characterization. In particular, if we make a boost from the AIE frame to the
heliocentric frame of our own measurements then the difference of the CMB
temperature dipole and the standard kinematic dipole identified
by (\ref{dipk}) can be observationally significant.

We will therefore define {\em general relativistic nonkinematic
differential expansion} (or more succinctly {\em relativistic differential
expansion}) to occur when the difference
\beq
\Delta T\Ns{nk-hel}={\Tmod\over\gmod(1-\bmod\cdot\nhel)}-{\TO\over\gfl(1-\bfl\cdot\nhel)}
\label{dT}
\eeq
has a measurably nonzero dipole when expanded in spherical
harmonics\footnote{By construction (\ref{dT}) has zero monopole.}, where
$\bmod=\vmod/c$ is the boost of the AIE frame in the heliocentric frame,
$\gmod=(1-\be\Ns{AIE}^2)^{-1/2}$, and
\beq\Tmod(\nLI)={T\Ns{CMB}\over 1+\zmod(\nLI)}\,,\label{Tani}
\eeq
is the anisotropic CMB temperature as measured in the AIE frame. Here
$T\Ns{CMB}=(1+z\ns{dec})\TO$ is the mean intrinsic temperature of the
primordial plasma at decoupling, $z\ns{dec}$ being the constant isotropic
redshift of decoupling in the FLRW model. In practice, ``measurably nonzero''
here means a contribution to (\ref{dT}) of at the least the same level,
$10^{-5}\TO$, as the primordial spectrum; i.e., one order of magnitude larger
than the boost dipole in (\ref{dipk}).

This an operational model--independent\footnote{The definition applies not only
to exact solutions of Einstein's equations, but to any cosmological model with
a close to linear Hubble law, including models with backreaction such as the
timescape model \cite{clocks,sol,obs,dnw}. One could further refine this
definition to treat the quadrupole and small frequency dependent effects of a
boost on the black body spectrum \cite{kk03}. However, as yet we are unable to
distinguish such terms given current knowledge of the modelling of foregrounds
such as galactic dust.} definition. Since eq.~(\ref{Tani}) does not separate
out the primordial CMB anisotropies, our definition cannot distinguish a
primordial CMB dipole from ``local'' nonkinematic relativistic effects. However,
we will construct numerical solutions constrained by large galaxy surveys,
leading to direct predictions for the amplitude and direction of the
nonkinematic dipole in terms of general relativity alone, rather than by
appealing to unknown physics in the early Universe.

Once one accounts for the known motions of the Sun and Milky Way within the
bound system that forms the Local Group of galaxies, to
explain the observed dipole in (\ref{dipk}) the LG has to move with a velocity
\beq \vo=635 \pm 38\kms\label{vLG}\eeq in the direction
\beq(\lcmb,\bcmb)=(276.4\deg,29.3\deg)\pm3.2\deg,\label{dcmb}\eeq
in galactic coordinates \cite{tsk08}. The results of \cite{rest,boost} show
that the CMB and AIE frames are statistically significantly different, while
the LG frame cannot be statistically distinguished from the AIE frame given
uncertainties in the data. Thus if we take the LG frame as the AIE
frame then (\ref{dT}) will involve the subtraction of two milli-Kelvin
anisotropies, one of which is nonkinematic, leaving a residual which is
likely to be observationally significant.

Our numerical simulations are not yet sophisticated enough that we will apply
(\ref{dT}) directly to sky maps. For the purpose of our numerical simulations
we will assume that the average isotropic expansion frame coincides with the
LG frame, so that $T\Ns{LG}=\Tmod$ and $\nLG=\nLI$ as given by (\ref{Tani}).
Furthermore, rather than working in the heliocentric frame, we will work in
the LG frame and constrain the CMB dipole and quadrupole of (\ref{Tani})
directly by ray tracing.

\subsection{Large scale homogeneous isotropic distance--redshift nonlinearity}
Independently of the energy momentum tensor, the luminosity distance relation
of any FLRW cosmology can be Taylor expanded at low redshifts to give
\beq d_L(z) = {c\over\HO}
\Bigg\{ z + {1\over2}\left[1-\qO\right] {z^2}
-{1\over6}\left[1-\qO-3{\qO\s}^2+\jO+ {kc^2\over{\HO\s}^2{\aO\s}^2}\right] z^3
 + O(z^4) \Bigg\}\label{flrw}
\eeq
where $d_L$ is the luminosity distance to the observed galaxy,
$\qO$ the deceleration parameter, $\jO$ the jerk parameter, $\aO=a(\tO)$
the present cosmic scale factor, and $k=-1,0,1$ the spatial curvature.
The $O(z^2)$ and higher order terms represent nonlinear corrections to the
linear Hubble law.

Even if the Universe is not described by a FLRW model, but by
some alternative in which a notion of statistical homogeneity applies, then
we can still expect a cosmic expansion law with a Taylor series in $z$
similar to (\ref{flrw}). This is the case for the timescape cosmology
\cite{clocks,sol,obs,dnw}, for example. However, in such cases the higher
order coefficients will not have the form given in (\ref{flrw}). In
particular, generically averages of inhomogeneous models do not expand
in such a way as to maintain a rigid constant spatial curvature, $k$.

\subsection{Small scale nonlinearities: the ``nonlinear regime''}
Even if the FLRW model is a good fit on large scales, the Taylor expansion
(\ref{flrw}) is only {\em a priori} justified on scales $d_L\gsim r\ns{ssh}$
on which a notion of statistical homogeneity applies. Small scale differences
will lead to complex deviations from an average linear Hubble expansion. In
the perturbed FLRW model such deviations are induced on small scales when
density perturbations become nonlinear, giving rise to the {\em nonlinear
regime} of perturbation theory.

In the standard cosmology small scale nonlinear cosmic expansion
is investigated with large $N$-body numerical simulations using Newtonian
gravity in a uniformly expanding box, with an expansion rate fit to a
FLRW model. While any form of nonlinear expansion might be
interpreted as ``differential expansion'', by construction the $N$--body
simulations can always be interpreted in terms of small scale
flows in which velocities are added with respect to the assumed FLRW
background. Such models do not allow the possibility of a
relativistic differential expansion as defined by eqs.\ (\ref{dT}),
(\ref{Tani}). The characterization of the differences between $N$-body
simulations and the models of this paper will be left to future work
\cite{next}.

In the present paper, we are interested in the observational scales which are
usually interpreted in the nonlinear regime of perturbed FLRW models, but we
make different model assumptions to interpret the observations. Regardless
of the assumed cosmological model, we use the term {\em nonlinear regime} to
apply to redshift--space distortions due to nonlinear expansion on scales of
tens of megaparsecs that affect all cosmological observations. For more
distant objects the redshift--space distortions at the source will lead to
small uncertainties as a fraction of the overall distance. However, in our own
vicinity on scales $d_L\lsim r\ns{ssh}$ these effects can be large.

In order to construct any numerical simulation some model is required.
In this paper, we will perform simulations with the Szekeres solution, which
does allow for a relativistic nonkinematic differential expansion. However, we
will first discuss the recent model--independent investigation of the
Hubble expansion in the nonlinear regime, which motivated the present
study.

\subsection{Model independent characterization of small scale nonlinear
expansion}

In recent work \cite{rest,boost} the problem of characterizing the Hubble
expansion below $r\ns{ssh}$ was approached
with no prior assumptions about the homogeneity of the spatial
geometry. In particular, given a large
data set with good sky coverage, one can simply determine the best fit average
linear Hubble law in radial shells, even in the regime in which the expansion
is nonlinear, a technique first used by
Li and Schwarz \cite{ls08}. Another alternative is to take angular averages, for example, by
employing a Gaussian window averaging method pioneered by
McClure and Dyer \cite{md07}.

Wiltshire \etal\ \cite{rest} applied these techniques to the \CS sample of 4,534 galaxies and
clusters compiled from earlier surveys by Watkins \etal\ \cite{wfh09,fwh10}. A startling
result was found --- when the best fit spherically averaged Hubble parameter
in inner shells was compared to the asymptotic value on $r>156\hm$ scales, it
was found that the Hubble expansion was more uniform in the rest frame of the
Local Group (LG) of galaxies than in the standard CMB
rest frame, with very strong Bayesian evidence. There is no reason why this
should be true in the standard cosmology. It is expected that the CMB
rest frame should
coincide with the local frame in which the Hubble expansion is most uniform,
with minimum statistical variations.

It was argued by Wiltshire \etal\ \cite{rest} that an arbitrary boost, $v$, of the central
observer from a rest frame in which the Hubble parameter, $H$, is close to
uniform will display a systematic offset of the value $H'$ determined by least
squares regression in spherical shells in the boosted frame. In particular, when
minimizing the sum $\chi_s^2=\sum_i\left[\si_i^{-1}(r_i-cz_i'/H')\right]^2$
with respect to $H'$, where $r_i$ and $\si_i$ are individual distances and
their uncertainties, under a boost of the central observer the original
redshifts transform as $cz_i\to cz_i'=cz_i+v\cos\phi_i$ for small $v$,
where $\phi_i$ is the angle between each data point and the boost direction.
Provided the number density of objects in a distance catalogue is balanced on
opposite sides of the sky, then terms linear in the boost cancel from
opposite sides of the sky in a spherical average, leaving a term proportional
to $v^2$. The offset is found to give approximately
\beq H'-H\simeq {v^2\over2\Ha\ave{r_i^2}},\label{offset}\eeq
in successive radial shells, where $\Ha$ is the asymptotic Hubble constant in
the range where expansion is linear. McKay and Wiltshire \cite{boost} found that such a signature
is indeed observed between the CMB and LG rest frames in both the \CS
sample \cite{wfh09,fwh10}, and in the larger {\em Cosmicflows-II} (CF2) sample
\cite{cf2}.

It was also found by Wiltshire \etal\ \cite{rest} that the largest residual
monopole variation in the Hubble expansion in the LG rest frame occurs
in a range $40\h$ -- $60\hm$, whereas the monopole variation in the CMB
frame is less than in the LG frame in this range only. Over the same
distance range $H'-H$ is found to deviate from the relation (\ref{offset})
in both the \CS and CF2 catalogues\footnote{The reported CF2 distances
include untreated distribution Malmquist biases \cite{cf2} which lead
to an additional spurious monopole \cite{HCT,boost} when spherical averages
are taken. Despite this bias the signature (\ref{offset}) is still apparent
in the CF2 catalogue in the {\em difference} $H'-H$, but with a somewhat
broader distance range, $30\h\lsim r\lsim67\hm$, over which (\ref{offset}) does
not apply, consistent with there being additional systematic uncertainties in
individual distances \cite{boost}.}. Angular averages
reveal a dipole structure in the Hubble expansion, whose amplitude changes markedly
over the range $32\h$ -- $62\hm$, in different ways in the two rest frames.
The conclusion from various analyses \cite{rest} is that the boost from the
LG frame to the CMB frame appears to be compensating for the effect on
cosmic expansion of inhomogeneous structures within this distance range.
A boost to the CMB frame has the effect of almost cancelling the monopole
and dipole variations; but not perfectly. Whereas the amplitude of the
dipole expansion variation declines to levels statistically consistent with
zero for $r\gsim65\hm$ in the LG frame, in the CMB frame the dipole amplitude
drops to a minimum value close to zero at $r\goesas44\hm$, but subsequently
increases \cite{rest}.

Finally, using Gaussian window averages, a sky map of angular Hubble expansion
variation on $r>15\hm$ scales was determined for the \CS sample and its
correlation coefficient, $\cal C$, with the residual CMB dipole in the LG frame
was computed. It was found that ${\cal C}=-0.92$ for an angular smoothing
scale $\sith=25\deg$, which was almost unchanged as the smoothing scale
was varied in the range $15\deg<\sith<40\deg$.

The combination of the above results led to the hypothesis of Wiltshire \etal\
\cite{rest} that a significant component of the observed CMB dipole, which is
conventionally attributed to a local boost (\ref{vLG}), (\ref{dcmb})
of the Local Group of galaxies, is nonkinematic in origin. It should be
attributed to a differential expansion of space, due to foreground
inhomogeneities on $\lsim65\hm$ scales which result in
a 0.5\% anisotropy in the distance--redshift relation below these scales.

From the point of view of general relativity, such a hypothesis is not
surprising --- it is simply a property of general inhomogeneous cosmological
models. Indeed, using a simple Newtonian approximation \cite{aa06} for LT
models \cite{L,T,B}, numerical estimates of the size of the effect were made
by Wiltshire \etal\ \cite{rest}. These were indeed consistent observationally both in
terms of the magnitude of the CMB temperature dipole and quadrupole,
and the scale of the void relative to that of the actual structures observed
in the nearby Universe \cite{el06}.

\section{The observational data}

\subsection{The anisotropy of the Hubble expansion}\label{HFani}

At low redshifts the Hubble constant for a spatially flat FLRW universe
can be calculated by rearranging the terms of (\ref{flrw}) to obtain
\beq
 \HO = \frac{c}{d_L} \left[z + \frac{1}{2}(1-\qO)z^2 - \frac{1}{6} (1-\qO-3{\qO\s}^2+\jO)z^3 \right].
\label{hubconst}
\eeq
Given a large set of data with independently measured values of $z$ and $d_L$
one could determine $\HO$, $\qO$ and $\jO$ for any spatially homogeneous
isotropic cosmology using the above formula.
In practice, with current data one is only able to independently determine $\HO$ at low redshifts, and when terms beyond the linear Hubble law are used in (\ref{hubconst}) then fixed values of $\qO$ and $\jO$ must be assumed from other observations. For example, the SH0ES \cite{shoes} estimate assumes values $\qO=-0.55$ and $\jO=1$ consistent with a spatially flat FLRW model with $\Omega_m=0.3$. In the present paper, we use $\qO=-0.5275$ and $\jO=1$, consistent with the best fit value $\Omega_m=0.315$ of Ade \etal\ \cite{Planck_params}.

\begin{figure}
\begin{center}
\includegraphics[scale=0.65]{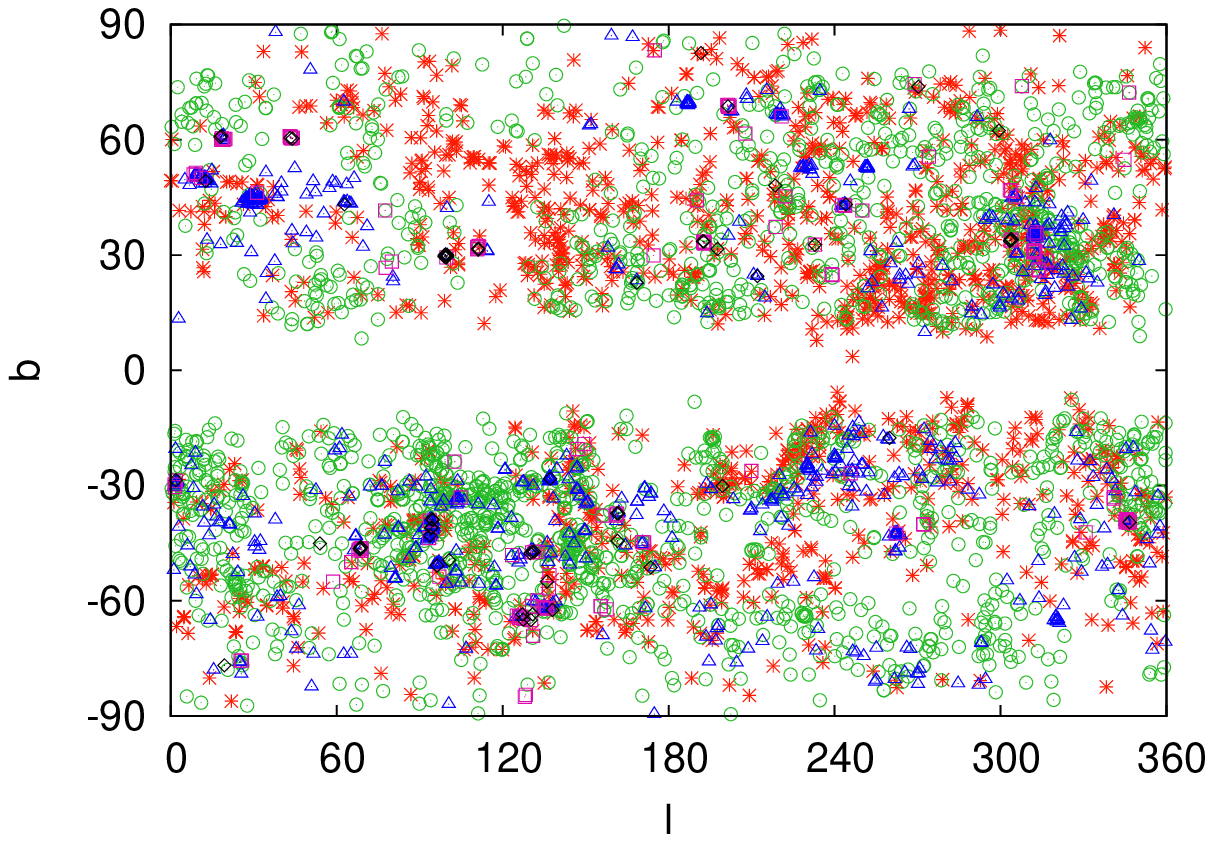}
\includegraphics[scale=0.65]{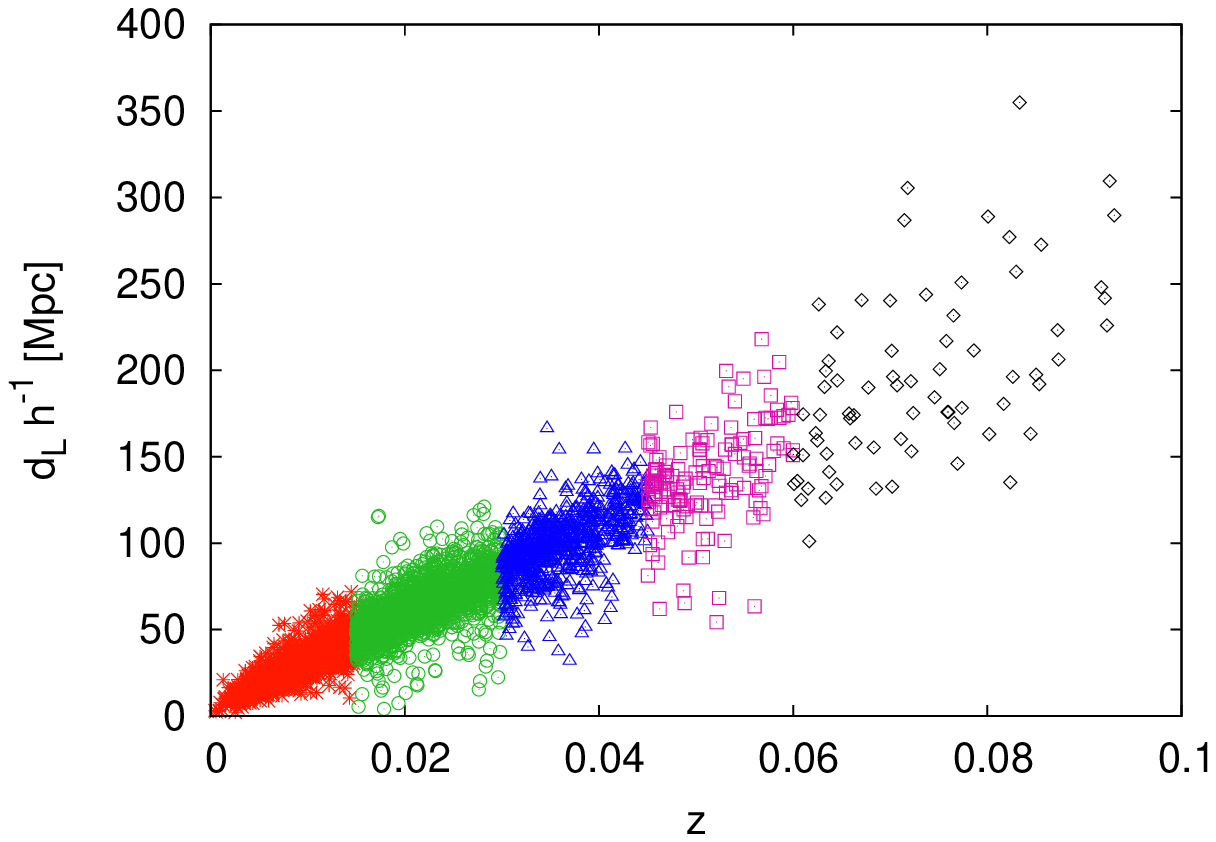}
\caption{The \CS sample. {\em Upper Panel}:
angular distribution of galaxies in the Galactic coordinates $\ell$ and $b$; {\em Lower Panel}: distance vs LG frame redshift;
$z \le 0.015$ red stars,
$0.015 < z \le 0.03$ green circles,
$0.03 < z \le 0.045$ blue triangle,
$0.045 < z \le 0.06$ magenta square,
$z > 0.06$ black diamonds.}
\label{COM-fig}
\end{center}
\end{figure}
As seen from Fig.\ \ref{COM-fig}, except for the Zone of Avoidance region obscured by our Galaxy ($|b|\lsim15^\circ$), the \CS sample has good angular coverage\footnote{This statement remains true when the data is broken into concentric radial shells in distance, as is seen in Fig.~2 of Wiltshire \etal\ \cite{rest}, where only the innermost of 11 radial shells (with $d_L<18.75\hm$) was found to have insufficient sky coverage when performing statistical checks.}, and thus can be used to evaluate large angle anisotropies of the Hubble expansion (dipole and quadrupole).
However, the \CS sample has large uncertainties associated with the distance measure.

From the point of view of propagation of uncertainty, it is better to work with the formula (\ref{flrw}) with $d_L$ as the independent variable in the numerator. To infer $\HO$ from the data we therefore
minimize the following sum
\beq \chi^2 = \sum_i \left( \frac{d_i - c \zeta_i /\HO} {\Delta d_i} \right)^2, \eeq
where
\beq \zeta_i = \left[ z_i + \frac{1}{2}(1-\qO)z_i^2 - \frac{1}{6} (1-\qO-3{\qO\s}^2+\jO)z_i^3 \right], \label{gami}\eeq
$d_i$ and $z_i$ are respectively the luminosity distance and redshift of each object in the \CS sample,
and $\Delta d_i$ is the distance uncertainty.
The above is equivalent to calculating the Hubble constant as a weighted average,
\beq\HO = {\sum_i H_i \wdi \over \sum_i \wdi},\label{Hs}\eeq where \beq H_i= c\zeta_i/d_i\label{Hi}\eeq
and
\beq \wdi= c\zeta_i d_i/(\Delta d_i)^2.\label{wd}\eeq
Wiltshire \etal\ \cite{rest} evaluated (\ref{Hs}) for spherical averages in independent radial shells, for the case of a linear Hubble law with $\zeta_i=z_i$, and separately considered angular averages using a Gaussian window function smoothing in solid angle.

Here we will apply Gaussian window function smoothing jointly in both solid angle and redshift, to obtain the following formula for the
average local Hubble constant centred at galactic coordinates ($\ell,b$) and redshift, $z$,
\beq
\HO(\ell,b,z) =
\frac{ \sum_i H_i \wdi \wzi \wti } { \sum_i \wdi \wzi \wti},
\label{HubEst}
\eeq
where $H_i$ and $\wdi$ are given by (\ref{Hi}) and (\ref{wd}) respectively, while
\beq \wzi = \frac{1}{\sqrt{2 \pi} \siz} \exp \left[- \frac{1}{2} \left( \frac{z-z_i}{\siz} \right)^2 \right], \eeq
\beq \wti = \frac{1}{\sqrt{2 \pi} \sith} \exp \left[- \frac{1}{2} \left( \frac{\thi}{\sith} \right)^2 \right],\label{wth}\eeq
$\siz = 0.01$, $\sith = 25^\circ$, and $\thi$ is the angle between the direction of each source $(\ell_i, b_i)$ and the direction of any given point on the sky, $(\ell,b)$:
\[ \cos \thi = \cos b \cos b_i \cos(\ell -\ell_i) + \sin b \sin b_i. \]

For $\wzi=1$ --- i.e., with no redshift smoothing --- equations (\ref{HubEst}) and (\ref{wth}) reduce to equations (B5) and (B9) derived in Appendix B of ref.\ \cite{rest} using a procedure based on minimizing the scatter in $H^{-1}$.

Using (\ref{HubEst}) we calculate the regional contributions to our locally measured Hubble constant on an
angular and redshift grid. For each redshift value on the grid, we construct
the angular maps of the Hubble expansion and express the Hubble flow in terms of
its fluctuations
\beq
\frac{\Delta \HO}{\av{\HO}} = \frac{\HO(\ell,b,z) - \av{\HO}}{\av{\HO}},
\label{dH}\eeq
where
\beq
\av{\HO} = \frac{\int {\rm d} \Omega ~ \HO(\ell,b,z)} {4\pi},
\eeq
is the spherically averaged value of (\ref{HubEst}).

For each redshift, the fluctuations (\ref{dH}) are then analysed using the spherical harmonic decomposition
\beq
\frac{\Delta \HO}{\HO} = \sum_{l,m} ~a_{lm} Y_{lm},\label{dHz}
\eeq
which allows us to evaluate the angular power spectrum:

\beq
C_l = \frac{1}{2l+1} \sum_m |{a}_{lm}|^2.
\label{cls}
\eeq

The power spectrum obtained in this way is subject to several biases and uncertainties \cite{HGNCPH2002}
\begin{equation}
C_l = \sum_{l'} M_{ll'} B_{l'}^2 {\cal C}_{l'} + N_{l'}
\label{clobs}
\end{equation}
where ${\cal C}_{l'}$ is the true underlying power spectrum, $M_{ll'}$ describes the mode--mode coupling resulting from incomplete sky coverage,
$B_l$ is a window function due to the smoothing, and $N_l$ is the noise.
As seen in Fig.\ \ref{COM-fig} for $|b|\lsim15^\circ$ data is incomplete in the galactic plane. In this paper, we do not mask these regions. Instead we extrapolate data to these regions using Gaussian smoothing of radius $\sith=25\deg$, as follows from eq.\ (\ref{clobs}). While this can potentially affect the inferred power spectrum, for the large angular scales (such as dipole and quadrupole) of interest here the results are not significantly altered \cite{rest}.
As for the noise, we estimate the level of contamination of the power spectrum due to distance uncertainties and number of data in the next section.

\subsection{Completeness and robustness}

As seen from Fig.\ \ref{COM-fig}, and in more detail in Fig.~2 of Wiltshire \etal\ \cite{rest}, there is good angular coverage in the \CS data. Potential systematic uncertainties from anisotropies generated by insufficient sky cover were investigated in detail by Wiltshire \etal\ \cite{rest}, who performed 12 million random reshuffles of the data in independent spherical shells, with the conclusion that for a binning scale $\Delta d=12.5\hm$ (or $\Delta z\simeq0.004$) results concerning the dipole anisotropy were robust on scales $0.002<z<0.04$, with up to 99.999\% confidence in some ranges.

In our case, we are also investigating the quadrupole anisotropy and adopt the larger redshift smoothing scale $\Delta z=0.01$.
However, there is still a possibility that the inferred anisotropy could result from some biases in the data.
To minimize any systematic bias and to confirm that the measured anisotropy is not spurious, we performed the following checks:
\begin{itemize}
\item We used the fluctuations (\ref{dHz})
rather than the spherical average (\ref{Hs}).
In the hypothetical case of an exactly homogeneous and isotropic universe (and perfect measurements) $\Delta \HO = 0$, so even if we have all the data only in one part of the sky and the rest of the sky without any measurement
we should not detect any anisotropy.

\item We shuffled the data.
We tested the robustness of the results on the dipole and quadrupole anisotropies by analyzing the reshuffled data --- for each pair of $z$
and $d_L$ we randomly reshuffled the angular position. We generated 100,000
reshuffled \CS catalogues and calculated the dipole and quadrupole of
the Hubble expansion. If the measured signal were comparable with the signal
obtained from reshuffled samples that would indicate that the original result
is spurious. That was not the case, however.

\item We used half of the data.
An alternative test of robustness was performed by taking
half of the \CS sample to calculate the dipole and quadrupole anisotropies
of the Hubble expansion. This was done for 100,000 randomly selected
halves of the original \CS catalogue. If the measured signal was
not consistent with the anisotropy obtained from half of the sample that
would indicate that the original result is spurious.
Again, this was not the case.
\end{itemize}

\begin{figure}
\begin{center}
\includegraphics[scale=0.65]{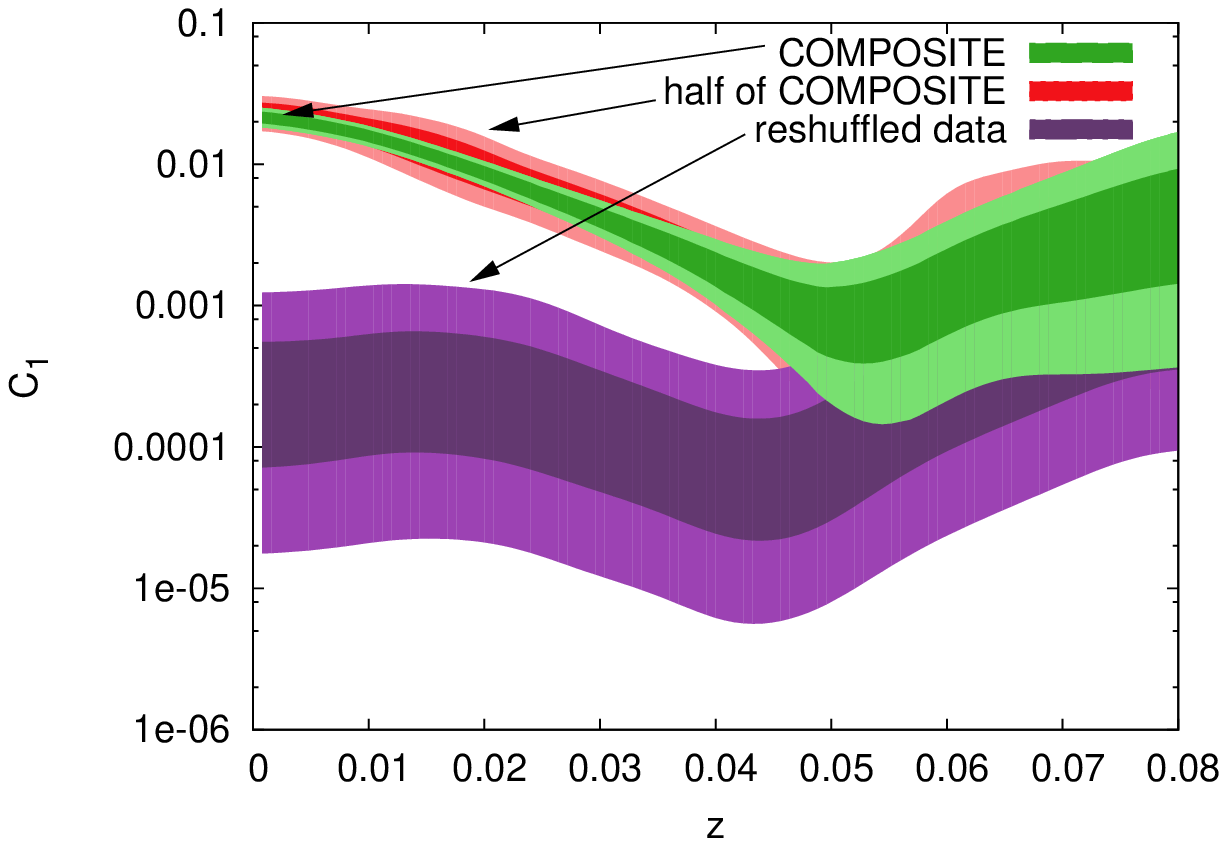}
\includegraphics[scale=0.65]{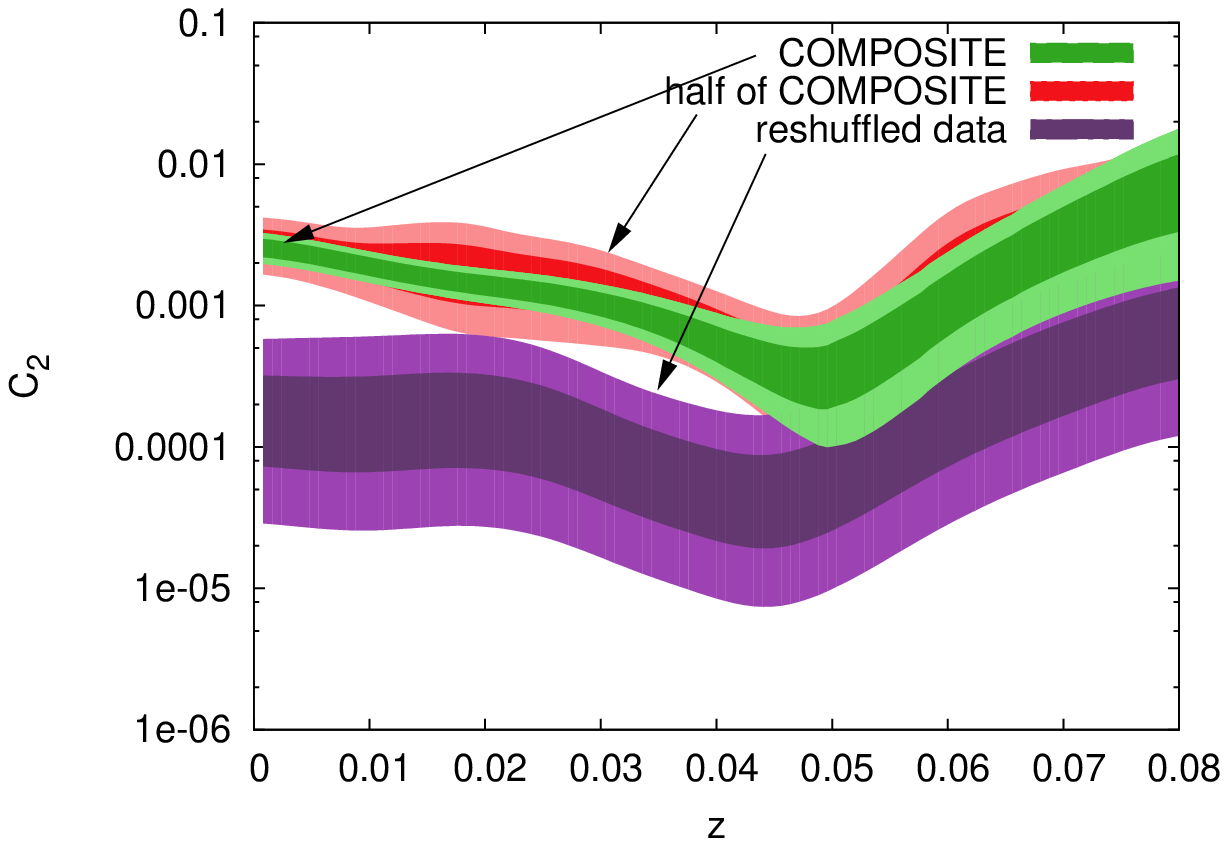}
\caption{The anisotropy of the Hubble expansion in the LG frame: dipole ({\em Upper Panel}) and quadrupole ({\em Lower Panel}). The green bands show the 65\% and 95\% confidence intervals for the \CS sample.
The red bands show the 65\% and 95\% confidence intervals obtained using 100,000 random halves of the \CS sample. The purple bands show the 65\% and 95\% confidence intervals obtained using 100,000 random reshuffles of the \CS sample.}
\label{COM-HFA}
\end{center}
\end{figure}

The results of the above analyses are combined in Fig.~\ref{COM-HFA}.
As seen our analysis passes these tests at the $2\si$ level for $z\lsim 0.045$.
This is consistent with the more exhaustive tests of Wiltshire \etal\ \cite{rest}, which showed that the dipole is not a systematic effect, to very high confidence.

\begin{figure}
\begin{center}
\includegraphics[scale=0.65]{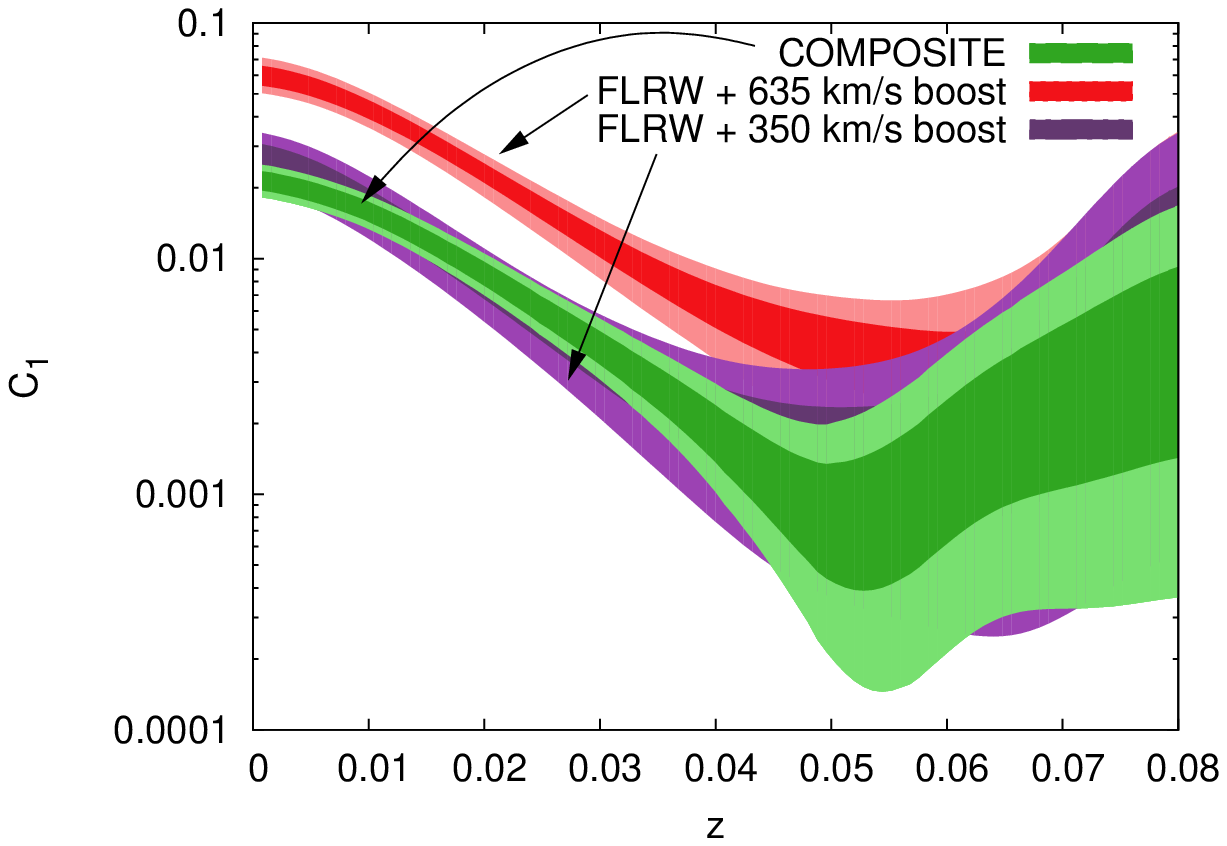}
\includegraphics[scale=0.65]{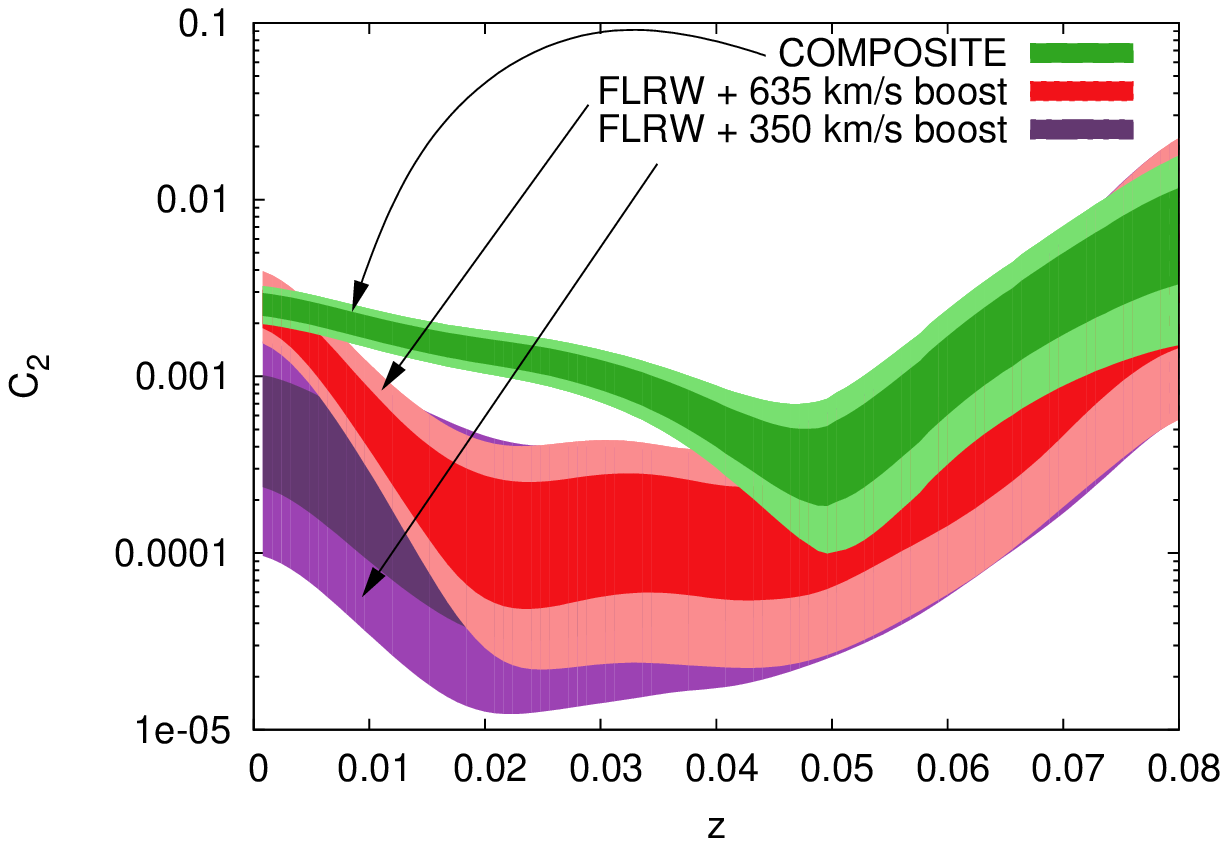}
\caption{The anisotropy of the Hubble expansion: dipole ({\em Upper Panel}) and quadrupole ({\em Lower Panel}). The green bands show the 65\% and 95\% confidence intervals for the \CS sample.
The red bands show the 65\% and 95\% confidence intervals obtained using mock \CS catalogues based on FLRW redshifts corrected by the local boost of $635$ km/s. The purple bands show the 65\% and 95\% confidence intervals for mock FLRW catalogues with a $350$ km/s boost in the same direction.}
\label{COM-boost}
\end{center}
\end{figure}
\subsection{Kinematic interpretation of anisotropies}\label{localboost}
The results presented in Fig.~\ref{COM-HFA} indicate the presence of anisotropy in the Hubble expansion up to $z\goesas0.045$, as determined from
the \CS sample redshifts transformed to the LG rest frame. The anisotropy is largest for small redshifts $z\sim0.02$, with the amplitude of the dipole dropping one order of magnitude from $z=0.02$ to $z=0.045$, from which point the dipole amplitude is consistent with that of the randomly reshuffled data at $2\,\si$.

According to the conventional explanation the anisotropy of the Hubble expansion observed in Fig.~\ref{COM-HFA} should have a kinematic origin, due to a boost (\ref{vLG}), (\ref{dcmb}) from the LG to CMB rest frame. This hypothesis can be directly tested by assuming a spatially homogeneous universe in the CMB frame, generating mock \CS samples in that frame, adjusting the redshift by performing a local boost to the LG frame, and then analysing the mock data in the manner of Fig.~\ref{COM-HFA}. Specifically,
\begin{enumerate}
\item We take the \CS sample. For each galaxy we have its angular position $(\ell_i,b_i)$, luminosity
distance $d_i$, uncertainty in distance $\Delta d_i$ and redshift $z_i$.
For each of these directions $(\ell_i,b_i)$ we use the FLRW
model to find the redshift ($z\Ns{FLRW}$) at which $d_L = d_i$, by solving
\[ d_L = (1+z) \frac{c}{\HO} \int\limits_0^{z\X{\rm FLRW}} {\rm d} z \frac{1}{\sqrt{\Omega_m(1 + z)^3 + 1 - \Omega_m}}\,,
\]
with $\Omega_m=0.315$, corresponding to the best fit parameters from the Planck satellite \cite{Planck_params}. (We also take the Planck satellite normalized $\HO=67.3\kmsMpc$. However, since we normalize all distances to $\hm$, this is inconsequential.)
\item
We adjust the redshift for the local boost \bea 1+z\Ns{FLRW-B}&=&\ga(1-\bo \cos \theta)(1+z\Ns{FLRW})\nonumber\\ &\simeq&(1- \bo \cos \theta)(1+z\Ns{FLRW}),\label{zboost}\eea where $\bo=\vo/c=(2.1\pm0.1)\times10^{-3}$ by (\ref{vLG}), while \[ \cos \theta = \cos \bcmb \cos b_i \cos( \lcmb - \ell_i) + \sin \bcmb \sin b_i, \] $\lcmb$ and $\bcmb$ being given by (\ref{dcmb}), and we have set $\ga=\left(1-{\bo\s}^2\right)^{-1/2}\simeq1$, ignoring terms of O$({\bo\s}^2)$.
\item
We construct a mock \CS catalogue in the LG frame, by replacing $z_i$ with $z\Ns{FLRW-B}$,
(i.e., the redshift obtained in the FLRW model adjusted for the LG motion).
\item
We construct 100,000 mock catalogues, using the known uncertainties. Firstly, each boosted redshift in step 3
is drawn randomly from a Gaussian distribution
taking into account the uncertainties in (\ref{vLG}) and (\ref{dcmb}).
Secondly, in step 1 each distance from
the \CS sample is replaced with a distance $d_{\cal N}$
\beq d_ {\cal N} = {\cal N} (\mu = d_i, \sigma=\Delta d_i) \label{dN}\eeq
equal to a random number drawn from a Gaussian distribution whose mean value
is $d_i$ with standard deviation equal to each individual distance uncertainty, $\Delta d_i$.
\item
For each of these mock catalogues we calculate the Hubble expansion and its anisotropy as outlined in Sec.~\ref{HFani}.
\end{enumerate}

The results are presented in Fig.~\ref{COM-boost}.
As seen the anisotropy produced by a FLRW model with local boost is characterized by a dipole three times larger than is observed in the \CS data at low redshift, and differs by more than 2$\si$ for all redshifts $z<0.04$.
On the other hand the quadrupole generated by the FLRW model with local boost is comparable to that of the \CS sample as $z\to0$, but becomes
smaller than that of the \CS sample for $z\gsim0.01$, being consistent
with that of the randomly reshuffled data in Fig.~\ref{COM-HFA}. Thus for $z>0.01$ the local boost of $635$ km/s cannot account for the amplitude of the observed quadrupole in the \CS sample.

We note that the linear $\cos\theta$ dependence in (\ref{zboost}) gives rise to a pure dipole anisotropy at fixed values of $z\Ns{FLRW}$ and $d_L$ in a linear Hubble relation. However, once (\ref{zboost}) is substituted in the Taylor series (\ref{gami}), a quadrupole and higher order multipoles are also generated at fixed redshift. The amplitude of the boosted quadrupole in Fig.~\ref{COM-boost} is larger than would be produced with perfect data, the ratio of the boost quadrupole and dipole contributions to (\ref{dHz}) generated by (\ref{gami}) and (\ref{zboost}) being proportional to ${\bo\s}^2\goesas4\times10^{-6}$. The relatively large value of the ratio $C_2/C_1\goesas0.05$ reflects the combination of the effect of angular smoothing in the Gaussian window average (\ref{HubEst})--(\ref{wth}) with the actual distance uncertainties assigned to the mock data, leading to $C_2\goesas0.004$ at low redshift for the randomly reshuffled \CS data in Fig.~\ref{COM-HFA}.

We have investigated by how much the magnitude of the local boost on the axis of the CMB and LG frames must be reduced in order to match the Hubble expansion dipole of the \CS sample. We find that a $350$ km/s boost would match the Hubble expansion dipole, giving results which are also shown in Fig.~\ref{COM-boost}. The quadrupole of the \CS sample is not matched, however. For a 350 km/s boost the quadrupole is consistent with the residual level of the randomly reshuffled data within 2$\sigma$ for all redshifts.

The interpretation of the anisotropy within a framework of the FLRW model
plus local boosts leads to a conundrum. The mismatch between the $350$ km/s
amplitude of a Local Group boost that would be consistent with the
Hubble dipole anisotropy and the $635$ km/s boost required to account for the
CMB dipole kinematically suggests two possible solutions:
(i) the galaxies in the \CS sample are in a coherent bulk flow with
respect to the CMB on scales up to $z\goesas0.045$; or (ii) the Hubble dipole
and other anisotropies contain a substantial nonkinematic component.

While the bulk flow hypothesis is the one that is widely studied --- being
based on the standard FLRW model --- it is
at odds with the results of \cite{rest} that the spherically averaged, or
monopole, Hubble expansion variation is very significantly reduced in the
LG frame as compared to the CMB frame on $\lsim70\hm$ scales. The spherical
average of a coherent bulk flow on such scales does not produce a monopole
expansion variation of the character seen in the \CS sample \cite{rest},
and such a result is not seen in $N$--body Newtonian simulations\footnote{The
effect of a local boost of the central observer is the most significant
aspect of our analysis. FLRW models with additional inhomogeneities produced
by Newtonian $N$--body simulations do not lead to results significantly
different from a pure FLRW model plus local boost shown as shown in
Fig.~\ref{COM-boost}. These results will be reported elsewhere \cite{next}.}.
Moreover, the signature of a systematic boost offset
(\ref{offset}) from the LG to CMB frame is seen in both the \CS and
Cosmicflows-2 samples \cite{boost}, providing a potential explanation
for the CMB frame monopole variation if the LG rest frame is closer to being
the frame in which anisotropies in the Hubble expansion are minimized.

We will now investigate the extent to which a nonkinematic interpretation
of the anisotropies is observationally consistent by ray tracing in
exact inhomogeneous solutions of the Einstein equations.

\section{Light propagation in the non-linear relativistic regime and the origin of anisotropies}

Relativistic cosmological models predict the expansion of the
Universe, which induces cosmological redshift. Cosmic expansion,
however, does depend on the local coupling of matter and curvature, and only in the FLRW model is expansion spatially homogeneous and isotropic.
The general relativistic formula for the redshift is \cite{E71}
\beq
\frac{1}{(1+z)^2} \frac{ {\rm d} z } { {\rm d} s } =
\frac{1}{3} \Theta + \Sigma_{ab} n^a n^b + u^a{}_{;b} u^b n_a,
\label{mr1}
\eeq
where $u^a$ is the matter velocity field,
$n_a$ is the connecting covector field locally orthogonal\footnote{I.e., $u^an_a=0$, $n_an^a=1$. In the case that the vorticity of the velocity field vanishes --- i.e., $u_{[a;b]}=0$ --- then $n_a$ is also the normal to a spatial hypersurface with tangent $u^a$. For practical purposes, this is taken to be the case in cosmological averages.} to $u^a$, $\Sigma_{ab}$ is the shear of the velocity field,
and $\Theta=u^a{}_{;a}$ is its expansion.
In the limit of spatially homogeneous and isotropic models the shear vanishes, $\Sigma_{ab} \to 0$, and
the expansion of the velocity field reduces to the Hubble parameter, $\Theta \to 3 H(t)$.
However, once cosmic structures form, the expansion field becomes non-uniform (ranging from $\Theta = 0$
inside virialized clusters of galaxies to $\Theta > 3 \HO$ within cosmic voids), and so the shear, $\Sigma_{ab}$, and acceleration, $u^a{}_{;b}u^b$, of the velocity field are nonzero.

Distances are also affected by presence of cosmic structures.
The general relativistic framework that allows us to calculate the distance is
based on the Sachs equations \cite{S61,P93}
\beq
\frac{{\rm d^2} d_A}{{\rm d} s^2} = - \left( \sigma^2 + \frac{1}{2} R_{ab}
k^a k^b\right) d_A, \label{e8}
\eeq
where $k^a$ is the tangent to null geodesics in a congruence,
$\sigma=\frac12\si_{ab}\si^{ab}$ is the scalar shear of the null geodesic
bundle, and $R_{ab}$ is the Ricci curvature.
The first term on the right hand side of (\ref{e8})
is often referred to as the Weyl focusing as it involves the Weyl
curvature,
while the second term is known
as the Ricci focusing. For the type of inhomogeneities considered in
this paper,
the amplitudes of the density contrast and density gradient are such that
the Weyl focusing is negligibly small compared to the Ricci focusing \cite{bf12}. Therefore, in this paper we work within the Ricci focusing regime\footnote{See ref.~\cite{bf12} for a detailed discussion on the applicability of Ricci focusing and the contribution of the Weyl curvature on light propagation. } and we neglect any contribution from $\sigma$. The luminosity
distance $d_L$ is then given by the reciprocity theorem \cite{E33,E71}
\beq
d_L = (1+z)^2 d_A. \label{etherington}
\eeq
Solving (\ref{e8}), (\ref{etherington}) for the areal and luminosity distances,
and (\ref{mr1}) for the redshift, we arrive at a general relativistic
distance--redshift relation, which will give rise to an anisotropic Hubble
expansion generated by the spatial inhomogeneities in the geometric terms on
the right hand sides of (\ref{mr1}) and (\ref{e8}). The anisotropies will
be most prominent over length scales characteristic of the matter
inhomogeneities, and will have characteristics which are distinct from
a simple FLRW geometry plus Lorentz boosts.

\subsection{The geometry and Einstein equations}

In order to solve (\ref{mr1}), (\ref{e8}) for the distance--redshift relation
we need to calculate all relevant physical quantities such as the Ricci
curvature, the shear of the null and timelike geodesic bundles, and the
expansion scalar.
For this purpose we use the Szekeres solution \cite{Sz}, which is the most general known exact solution of the Einstein equations for an inhomogeneous dust source. In the limit of a spatially homogeneous matter distribution it reduces to the FLRW model.

The advantage of the Szekeres model over the perturbed FLRW model\footnote{We
will present a comparison of the distinct differences from $N$--body
simulations in a future paper \cite{next}.} is that we can account for possible
nonkinematic differential expansion which we find to be associated with actual
observed structures in the local Universe. In particular, we will study a
quasispherical Szekeres model generated by a spherical void onto which an
additional inhomogeneity with an axial density gradient is superposed. Thus
we have both overdense and underdense regions in the same exact solution of
Einstein's equations.

The Szekeres model reduces to the spherically symmetric LT model in the limit of no superposed axial density gradient. In the LT limit the anisotropy in the Hubble expansion is generated solely by the off--centre position of an observer relative to the centre of the inhomogeneity. In the Szekeres model this parameter will still play a role. However, the angle between the observer and the density gradient axis will also give rise to more complex and realistic anisotropies than are possible with the LT model alone. This allows us greater freedom to more closely model actual structures in the local Universe. Of course, the Szekeres model still has limitations as to what it can describe. (We will return to this issue later on.)

The metric of the quasispherical Szekeres solution \cite{Sz,Szq} is usually represented in the
following form
\beq
\dd s^2 = c^2\dd t^2 - \frac{\left(R' - R \frac{{\cal E}'}{{\cal E}}\right)^2}
{1 - k}\dd r^2 - \frac{R^2}{{\cal E}^2} (\dd p^2 +\dd q^2), \label{ds2}
 \eeq
where ${}' \equiv \partial/\partial r$, $R = R(t,r)$, and $k = k(r)\leq1$ is
an arbitrary function of $r$. The function $\cal E$ is given by
 \beq
{\cal E}(r,p,q) = \frac{1}{2S}(p^2 + q^2) - \frac{P}{S} p - \frac{Q}{S} q + \frac{P^2}{2S} + \frac{Q^2}{2S} + \frac{S}{2} ,
 \eeq
where the functions $S = S(r)$, $P = P(r)$, $Q = Q(r)$, but are otherwise arbitrary. We take the coordinates $r$, $p$, $q$ and the functions $P$, $Q$, $R$, $S$, $\cal E$ all to have dimensions of length. We can also define angular coordinates, ($\theta$,$\phi$), by
\beq
p-P=S\cot\frac\theta2\cos\phi,\qquad
q-Q=S\cot\frac\theta2\sin\phi.
\eeq
Then ${\cal E}=S/(1-\cos\theta)$, and the metric (\ref{ds2}) takes the form
\bea
\dd s^2 &=& c^2\dd t^2 - \frac1{1 - k}\left[R'+
\frac RS\left(S'\cos\theta+N\sin\theta
\right)\right]^2\!\dd r^2
-\left[S'\sin\theta+N\left(1-\cos\theta\right)\over S\right]^2R^2\dd r^2
\nonumber\\ &&
-\,{\left[(\partial_\phi N)\left(1-\cos\theta\right)\over S\right]^2R^2dr^2}
+\,{2\left[S'\sin\theta+N
\left(1-\cos\theta\right)\right]\over S}\,R^2\dd r\,\dd\theta\nonumber\\ &&
-\,{2(\partial_\phi N)\sin\theta
\left(1-\cos\theta\right)\over S}\,R^2\dd r\,\dd\phi
-\,{R^2} (\dd\theta^2 + \sin^2\theta\,\dd\phi^2), \label{ds3}
 \eea
where $N(r,\phi)\equiv\left(P'\cos\phi+Q'\sin\phi\right)$.

The Einstein equations with cosmological constant, $\Lambda$, and
dust source of mass density, $\rho$,
\beq
G_{ab} - \Lambda g_{ab}=\kappa^2\rho\,u_au_b,
\eeq
where $\kappa^2= 8\pi G /c^4$, reduce to the evolution equation and the mass distribution equation.
The evolution equation is
\beq\label{evo}
\dot{R}^2 = -k(r)+\frac{2 M(r)}{R} +\frac 1 3\Lambda c^2 R^2,
\eeq
where $\dot{} \equiv \partial/\partial t$,
and $M(r)$ is a function related to the mass density by
\beq\label{rho}
\kappa \rho = \frac {2 \left(M' - 3 M {\cal E}' / {\cal E}\right)} {R^2
\left(R' - R {\cal E}' / {\cal E}\right)}.
\eeq
Note that
\beq\label{Ed}
{\cal E'\over E}={-1\over S}\left[S'\cos\theta+N\sin\theta\right]
\eeq
is the only term in (\ref{rho}) which gives a departure from spherical
symmetry.
One is free to specify the various functions as
long as (\ref{evo})--(\ref{Ed}) are satisfied.
Since $R(t,r)$ is the only function that depends on time,
(\ref{evo}) can be integrated to give
\beq
t - t\Z B(r) = \int\limits_0^{R}\frac{{\rm d} \widetilde{R}}{\sqrt{- k +
2M / \widetilde{R} + \frac 1 3 \Lambda c^2\widetilde{R}^2}},
\label{tbf}
\eeq
where $t\Z B(r)$ is one more arbitrary function called the {\it bang time} function, which describes the fact that the age of the Universe can be position dependent. If we demand that the age of the Universe is everywhere the same for comoving observers --- the homogeneous Big Bang condition --- then the above equations link $M(r)$ and $k(r)$. In the generic case $M$ and $k$ can be arbitrary, which could mean either a non-uniform Big Bang, or some turbulent initial conditions; i.e., conditions that would require a more complicated model than the Szekeres model.

\begin{figure*}[t]
\begin{flushleft}
\hbox{\hskip-15mm\includegraphics[scale=0.5]{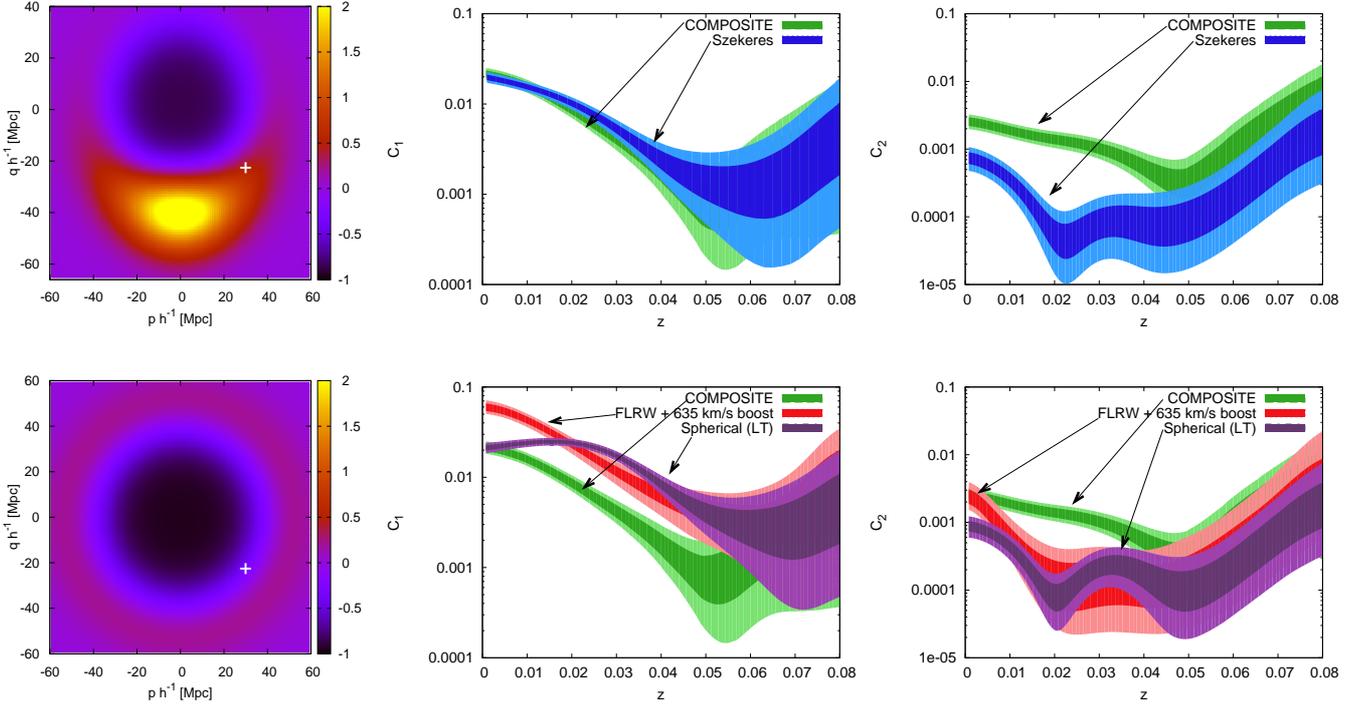}}
\caption{The anisotropy of the Hubble expansion. {\em Upper Panels} show the anisotropy of the Hubble expansion within the Szekeres model, whose density distribution is presented in the upper left panel (this panel shows the density contrast, $\de_{\rho}=(\rho-\bar\rho)/\bar\rho$ given in the right hand scale, when smoothed with Gaussian kernel of size $8\hm$); the position of the observer is marked with a cross ``+''. {\em Lower Panels} show the anisotropy of the Hubble expansion within the spherical model (LT) model, whose density distribution is presented in the lower left panel (cross ``+'' marks the position of the observer); the anisotropy of the Hubble expansion evaluated within the FLRW model with a local boost of 635 km/s is also presented for comparison. This shows how important matter anisotropies are to fully account for the observed anisotropies of the Hubble expansion.}
\label{Szek-hubf}
\end{flushleft}
\end{figure*}

The matter distribution in the Szekeres model has a structure of a dipole superposed on a monopole, (cf., upper left panel of Fig.~\ref{Szek-hubf}).
In order to determine the Szekeres model and solve all the equations, we need to specify its five arbitrary functions.
These are: $M$ and $k$ which describe the monopole distribution, and
$S$, $P$, and $Q$ which describe the dipole. If $S$, $P$, $Q$ are constant the dipole vanishes and we recover spherical symmetry; if $S'\ne 0$, and $P'=0=Q'$ then the model is axially symmetric.
These five functions (or any other combination of functions from which these can be evaluated) are sufficient
to solve all the equations that describe the evolution of matter and light propagation in the evolving geometry.

In the FLRW limit when the model becomes spatially homogeneous and isotropic
we have:
\begin{eqnarray}
R &\to& r a(t) \\
M &\to& M\Z0 r^3, \label{mass} \\
k &\to& k\Z0 r^2,\end{eqnarray}
where $M\Z0 = \frac{1}{2}{\HO\s}^2\,\Omega_m $,
$k\Z0 = {\HO\s}^2 (\Omega_m + \Omega_\Lambda-1)$,
and the functions $S$, $P$, and $Q$ are constant ($S' =0 = P' = Q'$).
Therefore, in the FLRW limit the dependence on $r$ in (\ref{evo}) cancels out and after dividing by $a^2$ we recover the well known form of the Friedmann equation
\[ H^2 = {\HO\s}^2 \left( \Omega_m a^{-3} + \Omega_k a^{-2} + \Omega_\Lambda \right),\]
where $\Omega_\Lambda = \Lambda c^2/(3{\HO\s}^2)$, and $\Omega_m + \Omega_k + \Omega_\Lambda =1$.

Let us then model the departure from homogeneity using the following profile of the mass function
\beq
M = M\Z0 r^3 \left[1 + \de_M(r)\right], \label{per1}
\eeq
where
\beq
\de\Z M(r) = \frac{1}{2} \deO \left( 1 - \tanh \frac{r-\rO}{2 \Delta r} \right),\label{Szekc1}
\eeq
with $-1\le\deO<0$, is a localized perturbation which is underdense at the origin. As $r\to\infty$, we have $\de\Z M\to0$ so that the spatial geometry is asymptotically that of the homogeneous and isotropic FLRW model. We normalize this geometry by choosing the spatially flat FLRW model which best fits the Planck satellite data, with $\Omega_m=0.315$ and $\HO=67.3\kmsMpc$ \cite{Planck_params}.

The function $k(r)$ is then evaluated from (\ref{tbf}) for each $r$ under the
assumptions:\break
(i) the age of the Universe is everywhere the same for comoving observers, $t\Z B = 0$; and\break (ii) $R(\tO,r)=r$ for each $r$, where the age of the Universe, $\tO$, is equal to that of the asymptotic background spatially flat FLRW model.

Finally we assume axial symmetry, with dipole described by only the function $S$, which we choose to be:
\begin{eqnarray}
&& S = r \left( \frac{r}{ 1 {\rm ~Mpc}} \right)^{\alpha-1}, \nonumber \\
&& P = 0, \nonumber \\
&& Q = 0,\label{Szekc2}
\end{eqnarray}
where $\alpha$ is a free parameter. When $\alpha \to 0$ the model becomes the spherically symmetric LT model, as shown in the lower left panel of Fig.~\ref{Szek-hubf}.

The model has 7 free parameters:
\begin{itemize}
\item 4 parameters that specify the Szekeres model
$\alpha$, $\deO$, $\rO$, $\Delta r$,
\item 3 parameters that specify the position of the observer
$\robs$, $\varphi\ns{obs}$, $\vartheta\ns{obs}$.
\end{itemize}
Since the model considered here is axially symmetric, we can choose the
observer to lie in the plane
$\varphi\ns{obs} = \pi/2 $ without loss of generality. In order to reduce the dimension of the parameter space we also set
$\Delta r = 0.1 \rO$
for simplicity. That leaves us with 5 parameters.
In reality, we expect the perturbations that describe actual cosmic structures to be much more complicated than the parameterization adopted here.
So while this gives us some flexibility, not all structures can be described using this parameterization.
The structures that can be described using this parameterization consist of a void and an adjacent overdensity,
as presented in the upper left panel of Fig.~\ref{Szek-hubf}. While not perfect, this parameterization aims to model
some of the major structures in the local Universe, such as the Local Void and the overdensity known as the Great Attractor \cite{tsk08}.

By varying the five free parameters, we can tune the size of the void and/or overdensity, the amplitude of the density contrast and the position of the observer relative to the structures.
We run a search through this 5-dimensional parameter space looking for a model
which as closely as possible satisfies constraints in the following order of importance:
\begin{enumerate}
\item The CMB temperature has a maximum value of $\TO+\Delta T$ relative to the mean $\TO=2.725\,$K, where
\beq \Delta T (\ell = 276.4^\circ, b = 29.3^\circ) = 5.77 \pm 0.36 ~{\rm mK},
\label{dcond}\eeq
which corresponds to the CMB temperature dipole amplitude and direction in the
LG rest frame.
\item The quadrupole of the CMB anisotropy is lower than the observed value \cite{pla}
\beq C\Ns{2,CMB}< 242.2^{+563.6}_{-140.1}\ {\rm \mu K}^2.\label{qcond}\eeq
While the dipole of the CMB is significantly affected by local expansion, the quadrupole is dominated by the baryonic physics of the early Universe, and the observed value itself is about 5 times smaller than the expectation based on the standard cosmology. Therefore we implement this constraint to ensure that the quadrupole generated by local inhomogeneities is much lower than the quadrupole generated at last scattering.

\item The dipole of the Hubble expansion anisotropy and its redshift dependence must be consistent with the observed anisotropy of the \CS sample as presented in Fig.~\ref{COM-boost}.

\item The quadrupole of the Hubble expansion anisotropy and its redshift dependence must be consistent with the observed anisotropy of the \CS sample as presented in Fig.~\ref{COM-boost}.
\end{enumerate}

\subsection{Constructing mock catalogues}\label{mockc}

The algorithm of our analysis can be summarized by the following steps:
\begin{enumerate}
\item
We first specify the Szekeres model.
\item
We apply the HEALPix grid of the sky and propagate light rays in these directions.
\item
We then calculate the CMB temperature maps.
\item
We use HEALPix routines to calculate the anisotropy of the CMB map.
\item
We take the \CS sample. For each galaxy we have its angular position $(\ell_i,b_i)$,
luminosity distance $d_i$, uncertainty in distance $\Delta d_i$ and redshift $z_i$.
For each of these directions $(\ell_i,b_i)$ we numerically propagate light rays,
using the null geodesic equations of the Szekeres model, up until $d = d_i$. We then write down the redshift evaluated within the Szekeres model $z\Ns{Sz}$.
\item
We construct the mock \CS catalogue, by replacing $z_i$ with $z\Ns{Sz}$, (i.e., the
redshift obtained in the Szekeres model for this direction and this distance).
\item
We construct 100,000 mock catalogues, by taking into account the actual
uncertainties, $\Delta d_i$, in the distances. As for the boosted FLRW mock catalogues, this is
done by replacing the distance from the \CS sample by $d_{\cal N}$ according
to (\ref{dN}).
\item
For each of these mock catalogues we calculate the Hubble expansion and its anisotropy as outlined in Sec.~\ref{HFani}.
\end{enumerate}

\subsection{Anisotropy of the Hubble expansion generated by cosmic structures modelled by the Szekeres model}\label{szekaniso}

Our search through the 5-dimensional parameter leads us to a model, whose mass profile as well as the position of the observer are presented in the upper left panel of Fig.~\ref{Szek-hubf}. The values of the free parameters are
\begin{eqnarray}
&& \alpha = 0.86, \nonumber \\
&& \deO = -0.86, \nonumber \\
&& \rO = 38.5~h^{-1} {\rm~Mpc}, \nonumber \\
&& \Delta r = 3.85~h^{-1} {\rm~Mpc},\label{Szekc3}
\end{eqnarray}
and the position of the observer:
\begin{eqnarray}
&& \robs = 25 h^{-1} {\rm~Mpc}, \nonumber \\
&& \varphi\ns{obs} = 0.5 \pi, \nonumber \\
&& \vartheta\ns{obs} = 0.705 \pi.\label{Szekc4}
\end{eqnarray}
These distances are coordinate distances, not radial proper distances or
luminosity distances.

Interestingly, if we take the region of maximum
overdensity in Fig.~\ref{Szek-hubf}, with density contrast $\de\rho/\rho>2$,
this region is found to be located at redshifts and luminosity distances in the
ranges $0.003\lsim z\lsim0.013$ and $16\h\lsim D_L\lsim53\hm$, comparable to those of the Centaurus cluster / Great Attractor\footnote{The Centaurus cluster, at LG frame redshift of $z=0.0104\pm0.0001$ \cite{sr99}, lies in the nearer portion of the Great Attractor region \cite{cpthc13}.}. We did not supply the redshift of the overdensity as an {\em a priori} constraint, but arrived at it using a grid search on possible Szekeres models following the criteria specified above.

Furthermore, in terms of the angular extent, the region with $\de\rho/\rho>2$ occupies an ellipsoidal region with $220\deg<\ell<320\deg$, and $-60\deg<b<40\deg$. The Centaurus cluster at $(\ell,b)=(302.4\deg,21.6\deg)$ on the near side
of the Great Attractor is thus contained with the overdense region. However,
the Norma cluster at $(\ell,b)=(325.3\deg,-7.3\deg)$ which is in the angular
centre of the Great Attractor, but on the far side in distance at $z=0.0141\pm0.0002$
in the LG frame falls $5\deg$ outside our overdense region by angle and
$\Delta z=0.0011$ by redshift. The fact that the alignment of the overdensity
is very close to the actual Great Attractor, but does not yet match precisely
is consistent with the fact that there are further features of the Hubble
expansion that we have still to account for -- its quadrupole -- as we shall
discuss below.

The anisotropy of the Hubble expansion within such a model is presented in the upper panels of Fig.~\ref{Szek-hubf}.
We find that the first three criteria given in Sec.~\ref{mockc} are all satisfied. In particular, the CMB temperature dipole is \[\Delta T\Ns{CMB} = 5.58 {\rm ~mK}, \] as is expected in the LG frame, while the quadrupole of the CMB temperature anisotropy is \[ C\Ns{2,CMB} = 8.26 {\rm~ \mu K}^2. \] Furthermore, the dipole of the Hubble expansion within this model is consistent with the dipole of the Hubble expansion inferred from the \CS sample.

The results shown in Fig.~\ref{Szek-hubf} display the clear advantage of the Szekeres model in comparison with a FLRW model plus local boosts, as discussed in Sec.~\ref{localboost}.
Applying a local boost to an otherwise homogeneous and isotropic universe we
found that it was not possible to fit the Hubble expansion dipole and the CMB temperature dipole simultaneously.
By contrast the Szekeres model simultaneously matches both the CMB temperature dipole in the LG frame and the Hubble expansion dipole of the \CS sample over all redshifts in the survey with sufficient data.
Given the fact that the structure of the Universe on scales below $100\hm$ is
very inhomogeneous, and that differential cosmic expansion is a generic feature of cosmological solutions of Einstein's equation, it should perhaps not be a surprise that the inhomogeneous model performs better.

On the other hand, the particular Szekeres model considered here is not able to reproduce the quadrupole of the Hubble expansion seen in the \CS sample, which is about three times larger in magnitude than in the simulation. The Hubble expansion quadrupole in the Szekeres model (\ref{per1})--(\ref{Szekc4}) in fact has an amplitude consistent with that of the randomly reshuffled data in Fig.~\ref{COM-HFA}, and is not statistically significant.

The fact that we can account for the Hubble expansion dipole, but not the quadrupole may well be due to the simplicity of the model (\ref{per1})--(\ref{Szekc4}), (cf., upper panel of Fig.~\ref{Szek-hubf}). In particular, the choice (\ref{Szekc2}) enforces an axial symmetry on the mass distribution, which could be altered to give finer details. This would require a more complex model, and is left for future investigations.

As an indication of how the properties of the Hubble expansion variation are induced by changes in the matter distribution, we have also investigated the anisotropy of the Hubble expansion evaluated using a spherical void LT model, which is obtained from the Szekeres model in the limit of a vanishing matter dipole, $\alpha \to 0$.

We use the same parameterization and procedure as outline above with $\alpha = 0$, which ensures spherical symmetry.
As in the case of a simple boost (Sec.~\ref{localboost})
we are not able to simultaneously fit the CMB temperature variation and the full redshift dependence of the Hubble expansion anisotropy. At best, we can only reproduce some of the features.

An example of this investigation is presented in lower panels of Fig.~\ref{Szek-hubf}.
The values of the free parameters are
\begin{eqnarray}
&& \alpha = 0, \nonumber \\
&& \deO = -0.95, \nonumber \\
&& \rO = 45.5~h^{-1} {\rm~Mpc}, \nonumber \\
&& \Delta r = 4.55~h^{-1} {\rm~Mpc}
\end{eqnarray}
and the position of the observer is
\begin{eqnarray}
&& \robs = 28 h^{-1} {\rm~Mpc}, \nonumber \\
&& \varphi\ns{obs} = 0.5 \pi, \nonumber \\
&& \vartheta\ns{obs} = 0.5 \pi.
\end{eqnarray}

The model matches the correct temperature dipole of the CMB in the LG frame
\[\Delta T\Ns{CMB} = 5.63 {\rm ~mK} \]
and the quadrupole of the CMB temperature anisotropy is
\[ C\Ns{2,CMB} = 20.73 {\rm~ \mu K}^2. \]
However, the Hubble expansion dipole anisotropy can only be matched at very low redshifts (see middle panel of Fig.~\ref{Szek-hubf}).
As the redshift is increased the magnitude of the dipole increases until for, $z>0.015$, it becomes consistent with that predicted by the FLRW model plus a local boost of $635$ km/s in the LG frame --- which was not consistent with the \CS data, however.

This illustrates the fact that the amplitudes of the CMB dipole and higher multipoles in LT models can be roughly estimated for off--centre observers by an effective Newtonian approximation \cite{aa06} using the velocity appropriate to a boosted observer in the FLRW geometry. This limit was discussed by Wiltshire \etal\ \cite{rest}, who gave an example of a LT void with a somewhat different mass profile but with similar parameters, being 18\% larger but with a less sharp density gradient.

An examination of the density profile panels of Fig.~\ref{Szek-hubf} illustrates the role that is played by differential cosmic expansion. In particular, Lorentz boosts represent a point symmetry in the tangent space of any general observer. In the case of the LT model, the axis which joins the centre of the void to the position of the off--centre observer defines a direction along which a radial boost can be taken to act from the void centre. Since the differential expansion is purely radial with respect to the centre, it still somewhat mimics the action of a point symmetry. Thus it is not surprising that the LT dipole becomes equivalent to the FLRW model plus boost on scales larger than $\rO+\robs$. By contrast, the Szekeres model incorporates a mass dipole on an axis distinct from that joining $\rO$ to $\robs$. This distributed density gradient therefore gives rise to a differential cosmic expansion which cannot be mimicked by a boost or any other point symmetry relative to the central point, $\rO$.

The models considered in this Section show how the presence of cosmic structures affect the anisotropy of the Hubble expansion. The more structures that are present in the model, the better is the consistency with observational data.

\section{Potential impact on CMB anomalies}\label{anomalies}

Any model cosmology in which (\ref{dT}) is nonzero will demand a different
to standard approach to the analysis of large angle CMB anisotropies. The
multipole expansion of (\ref{dT}) will consist of terms which could be deemed
to be ``anomalous multipoles'' relative to the kinematic expectation. As the
dipoles will not cancel perfectly, to leading order there will generally be
an ``anomalous dipole'' which may demand a reexamination of the observed power
asymmetry and related large angle anomalies \cite{Piso,Piso2}. This would have
a major impact on observational cosmology, as has already been discussed
by Wiltshire \etal\ \cite{rest}.

While one should naturally be sceptical of any suggestion that large angle CMB
anomalies result from a nonkinematic relativistic differential expansion
on $\lsim70\hm$ scales, a very important implication of the present paper
is that ray-traced exact solutions similar to those described here will
provide, for the first time ever, concrete models that can actually be
tested against Planck satellite data for their effect on large angle
anomalies. Such a project might even demand subtle changes in the
treatment of the galactic foreground in the map making procedures. Thus it is
important to first have the best possible model of relativistic
differential expansion before embarking on such a challenge.

A detailed analysis of the multipoles of (\ref{dT}) is therefore left
for future work. In particular, we need to first refine the
Szekeres model to incorporate additional structures giving a Hubble
expansion quadrupole with the observed redshift dependence. Since the CMB
quadrupole for the Szekeres model (\ref{Szekc3}), (\ref{Szekc4}) is 30 times
smaller in amplitude than the observed CMB quadrupole, we should reasonably
expect that this can be accommodated. The redshift dependence of the Hubble
expansion dipole and quadrupole seen in the \CS sample should also to be
confirmed with other data sets\footnote{In the case of the Cosmicflows-2
sample \cite{cf2}, for example, this requires a careful treatment of Malmquist
biases to remove a monopole bias \cite{HCT,boost}.}.

In this paper, we have considered LT and Szekeres models in the the LG frame
treated as the average isotropic expansion frame, with a ray-traced CMB
treated according to (\ref{Tani}), as this is computationally simpler. With a
refined Szekeres model one should also boost to the heliocentric frame and
constrain the ray traced simulations with the complete sky map of the
observed heliocentric dipole from Planck satellite data with an appropriate
galactic sky mask applied, rather than adopting our simpler procedure
(\ref{dcond}) of just matching the amplitude of the equivalent temperature
dipole in the LG frame.

While one cannot know the outcome of any such simulations before performing
them, there are as yet no obstacles to the possibility that such
investigations will result in observationally consistent alternative models
of the large angle CMB sky. In particular, as was discussed in ref.\
\cite{rest}, the claim of the Planck team \cite{Pboost} that the
kinematic nature of the transformation from the heliocentric to CMB frames
has been verified by the effects of frequency modulation and aberration in
the CMB anisotropy spectrum actually depends on angular scale. The boost
direction coincides with the expected direction $(\ell,b)=(264\deg,48\deg)$
only for small angle multipoles $l\ns{min}=500<l<l\ns{max}=2000$. For large
angle multipoles $l<l\ns{max}=100$ the inferred boost direction moves across
the sky to coincide with the modulation dipole anomaly direction \cite{heb09},
$(\ell,b)=(224\deg,-22\deg)\pm24\deg$. Since the nonkinematic terms in
(\ref{dT}) will only affect large angle power, this angular scale dependence
of the results of \cite{Pboost} and their association with the anomaly
direction is perhaps suggestive.

\section{Conclusion}

Cosmic structures such as voids, sheets, filaments, and knots participate differently in the expansion of the Universe.
The expansion rate gradually changes from no expansion inside virialized high density regions such as superclusters to
a higher than average expansion rate inside voids. This differential expansion of the space can be observed in the anisotropy of the Hubble expansion, especially on scales up to a few hundreds megaparsecs. In general relativity, differential cosmic expansion is the norm in all cosmological models which are not spatially homogeneous and isotropic. The anisotropy of the Hubble expansion is thus expected to quantitatively differ from that of a FLRW model in which all departures from homogeneity can be described by local Lorentz boosts of the source and observer.

The effects that we consider in this paper appear to have been largely
overlooked as serious possibilities in the past for two reasons. Firstly,
in considering nonlinear anisotropies many cosmologists typically think about
the Rees--Sciama effect \cite{RS}, in which one considers a photon traversing
{\em from one average position across a nonlinear structure to another average
position}. Such considerations miss the peculiar potential effect of placing
observers deep inside the nonlinear structures (cf.\ Fig. 6 in ref.~\cite{B09}).
When we take the same structures studied in this paper and place observers
far from the structures then the amplitude of the temperature anisotropies
is of order $|\Delta T|/T<3\times10^{-7}$ consistent with previous
estimates which use larger voids and generate a somewhat larger
amplitude \cite{RRS,is06}.

Secondly, simple order of magnitude estimates suggest that a Rees--Sciama
dipole will in general generate a Rees--Sciama quadrupole of similar order
\cite{MM}, and one might na\"{\i}vely assume that similar arguments apply
to all general nonlinear distance--redshift anisotropies. However, our
results show again that one cannot extrapolate the argument for the
Rees--Sciama effect involving both a source and observer far from a structure
to the case of an observer inside the structure. While it is certainly
possible that the relative size of the CMB quadrupole and dipole would be
comparable at certain locations in other structures, we find that for
observers placed at any position in the Szekeres model (\ref{Szekc3}),
(\ref{Szekc3}) the quadrupole is always much smaller than the bound
(\ref{qcond}). Thus when one is dealing with observers inside a
nonlinear structure the details of the density profile and the observer's
position are crucially important. In our case,
we have a {\em particular location} relative to structures such as
the Local Void and the ``Great Attractor''. Our study is the first to
benefit from constraining ray tracing simulations with actual large galaxy
surveys outside a framework of the FLRW cosmology plus local boosts.

In this paper we investigated the anisotropy pattern of the Hubble expansion, considering the dipole and quadrupole variations in the LG frame. Most previous studies have either focused on the monopole, i.e., the global (average) value of $\HO$, or on bulk flows. In a way this is analogous to studies of the CMB in 1970s--1990s. However, with increasing amount of data and precision of measurements, we are slowly arriving at the stage where we can study anisotropies of the Hubble expansion, just as we now study the anisotropy of the CMB temperature fluctuations.

In analogy to CMB temperature fluctuations, we show that the Hubble expansion can be decomposed using spherical harmonics and expressed in terms of an angular power spectrum.
Moreover, by averaging data at various redshifts we can have additional information about the redshift dependence of the multipoles of the Hubble expansion.
Irrespective of any theoretical assumptions about cosmic expansion, this is a novel technique that carries complimentary if not additional information to studies of bulk flow that have been extensively carried out in the past years.

In Sec.~\ref{HFani} we developed the formalism used to study the anisotropy of the Hubble expansion. When applied to the \CS sample we identified the presence of dipole and quadrupole anisotropies in the Hubble expansion. These anisotropies are statistically significant in the data up to $z\lsim0.045$. For larger redshifts the amount of data is small and the signal is no longer distinguishable from noise.

We compared the measured anisotropy with predictions from a FLRW model assumed to be homogeneous and isotropic in the CMB frame, and also particular LT and Szekeres models with small scale inhomogeneities in the LG frame. All models were assumed to be identical to a spatially flat FLRW models on scales $\gsim100\hm$, with parameters fixed to those of the FLRW model that best fits the Planck satellite data \cite{Planck_params}. The FLRW model with a local boost from the CMB to LG frame did not fit the observed redshift dependence of the dipole of the Hubble expansion of the \CS sample as seen in the LG frame. In order to match the observed features of the dipole of the Hubble expansion, the local boost would have to be reduced to approximately $350$ km/s, which is much smaller than the actual $635 \pm 38$ km/s that is required if the CMB temperature dipole is purely kinematic.

A quasispherical Szekeres solution that allows for variations of the local geometry generated by the presence of cosmic structures, which effectively model the Local Void and ``Great Attractor'', was found to improve the fit. This analysis shows that the local cosmological environment does affect the Hubble expansion. Physically, this can be understood in terms of the differential expansion of the space, with the void expanding faster and the overdensity expanding at a slower than the average expansion rate.

As yet, the numerical model does not have a Hubble quadrupole as large as that seen in the \CS sample. However, if extra modifications are added --- for example, by using methods to include extra structures \cite{SG15b} --- then given the magnitude of the effects that remain to be explained, it is highly plausible that highly accurate models of the local cosmic expansion can be developed.

All our models are constrained by a match to the magnitude and direction of the
CMB temperature dipole. Since the models are nonlinear the addition of further
structures can affect the alignment and scale of the structures, and the
position of the observer, as compared to a simpler model. In moving from the simple
LT void model to our single void / single overdensity Szekeres model, for
example, the scale of the void was reduced by 40\%, while also achieving
a fit to the Hubble expansion dipole over a range of redshifts.

The effect which remains to be explained -- the Hubble expansion quadrupole
-- is an order of magnitude smaller than the Hubble expansion dipole.
Therefore we should not expect such large changes of scale as occurred between
the LT and Szekeres models already studied. However, we note that the overdensity in our simple Szekeres model overlaps with the observed Great Attractor in both angle and redshift on the near side but not completely on the far side.
Furthermore, the additional major structures that should still be accounted
for include most notably the Perseus--Pisces concentration, which lies at
LG frame redshifts $0.0182$--$0.0194$. This is at the upper end of the
redshift/distance range of the structures that we are considering, with a
a likely impacting on the alignment of the far side of the overdensity which
we have identified with the Great Attractor. Whether this
can be done while also accounting for the Hubble expansion quadrupole
is an important question left for future work.

As discussed in Sec.~\ref{anomalies}, our approach may potentially provide a simple physical explanation of particular large anomalies in the CMB radiation, in terms of known physics. But this is a matter for future investigations.

The main result of this paper is that with just the FLRW geometry plus a local boost of the Local Group of galaxies it is impossible to simultaneously fit both the CMB dipole and quadrupole anisotropies and the redshift dependence of the dipole anisotropy of the local expansion of the Universe, determined by the \CS\ sample. To explain the observed features we need to use models that exhibit differential cosmic expansion. Further refinement of such models may potentially have a major impact on cosmology.

\acknowledgments We thank Fran\c{c}ois Bouchet, Thomas Buchert, Lawrence Dam, Syksy R\"as\"anen, Nezihe Uzun and
Jim Zibin for helpful discussions and correspondence. This work was supported by
the Marsden Fund of the Royal Society of New Zealand, and by the Australian
Research Council through the Future Fellowship FT140101270. Computational
resources used in this work were provided by Intersect Australia Ltd and the
University of Sydney Faculty of Science.

\end{document}